\documentclass[a4paper,USenglish]{dagrep}

\usepackage{microtype}

\usepackage{todonotes}
\usepackage{tabularx}
\usepackage{booktabs}

\hypersetup{
	colorlinks=true,
	linkcolor=black,
	citecolor=black,
	filecolor=black,
	urlcolor=black,
	pdfauthor={Andre van Hoorn, Pooyan Jamshidi, Philipp Leitner, Ingo Weber},
	pdftitle={Report from GI-Dagstuhl Seminar 16394: Software Performance Engineering in the DevOps World},
}

\subject{Report from GI-Dagstuhl Seminar 16394}
\title{Software Performance Engineering in the DevOps World}
\titlerunning{16394 --- Software Performance Engineering in the DevOps World}

\author[1]{Andre van Hoorn}
\author[2]{Pooyan Jamshidi}
\author[3]{\newline Philipp Leitner}
\author[4]{Ingo Weber}
\affil[1]{University of Stuttgart, DE
}
\affil[2]{Imperial College London, GB}
\affil[3]{University of Zurich, CH}
\affil[4]{Data61, CSIRO, AU}
\authorrunning{Andre van Hoorn, Pooyan Jamshidi, Philipp Leitner, Ingo Weber}

\seminarnumber{16394}
\semdata{September 25--30, 2016 --- \url{http://www.dagstuhl.de/16394}}

\begin{document}

\maketitle

\begin{abstract}
This report documents the program and the outcomes of GI-Dagstuhl Seminar 16394 ``Software Performance Engineering in the DevOps World''. %
\end{abstract}

\section{Executive Summary}
\summaryauthor[Andre van Hoorn, %
Pooyan Jamshidi, %
Philipp Leitner, and %
Ingo Weber]{%
Andre van Hoorn (University of Stuttgart, DE)\\ %
Pooyan Jamshidi (Imperial College London, GB)\\ %
Philipp Leitner (University of Zurich, CH)\\ %
Ingo Weber (Data61, CSIRO, AU)
}

\license



The seminar addressed the problem of performance-aware DevOps. Both, DevOps and performance engineering have been growing trends over the past one to two years, in no small part due to the rise in importance of identifying performance anomalies in the operations (Ops) of cloud and big data systems and feeding these back to the development (Dev). However, so far, the research community has treated software engineering, performance engineering, and cloud computing mostly as individual research areas. We aimed to identify cross-community collaboration, and to set the path for long-lasting collaborations towards performance-aware DevOps.

The main goal of the seminar was to bring together young researchers (PhD students in a later stage of their PhD, as well as PostDocs or Junior Professors) in the areas of (i) software engineering, (ii) performance engineering, and (iii) cloud computing and big data to present their current research projects, to exchange experience and expertise, to discuss research challenges, and to develop ideas for future collaborations.

\tableofcontents


\section{Overview of Talks}

\abstracttitle{Performance Engineering for DevOps Using Survivability Modeling of High Availability Systems (Keynote)}
\abstractauthor[Alberto Avritzer]{Alberto Avritzer (Sonatype, USA, aavritzer@sonatype.com)}

\license


As our society evolves, more and more aspects of our daily life depend on large-scale infrastructures such as computer infrastructures, rails and road networks, gas networks, water networks, power networks, and telecommunication networks, including the Internet, wired and wireless telephony. Critical infrastructures are everywhere and they are becoming increasingly more interconnected and interdependent.
As networks become smarter, they rely more heavily on information and communication technologies, also known as ICT. For this reasons, a failure in the ICT network can cause problems in different critical infrastructures. As another example, a failure in the power network can cause disruptions in a number of different networks.
We present a Nexus Sonatype based DevOps approach for continuous performance improvement of high-availability systems that is composed of several important components: i) large repositories representing the customer environment, ii) automated analysis of performance tests, iii) JMeter load testing scripts and associated data.

We  present high-availability models of critical infrastructures and we connect our DevOps continuous performance improvement process with our approach to  the system availability assessment, e.g., through reduced failure recovery times.


\abstracttitle{DevOps---How a Fortune 500 Company Translates Theory in Reality (Keynote)}
\abstractauthor[Oliver Beck]{Oliver Beck (SAP, Germany, oliver.beck@sap.com)}

\licenseIngo

In this talk, Oliver Beck presented SAP's approach to managing DevOps at scale, and how they address practical challenges of changing culture in a large, well-established organization. Oliver Beck is a Vice President at SAP, where he is responsible for global DevOps initiatives, among others. (Text written by Ingo Weber)


\abstracttitle{Performance Regression Analysis in the DevOps World}
\abstractauthor[Cor-Paul Bezemer]{Cor-Paul Bezemer (Queen's University, Canada, cpbezemer@gmail.com)}

\license

A performance regression occurs when an application update unintendedly slows down the application. For example, the CPU usage or execution time of a task increases considerably after deploying an update. The goal of performance regression analysis (PRA) is to detect such performance regressions and analyze what causes them.

Most approaches for PRA follow a similar process to detect and analyze performance regressions. For every application update, a performance test is executed while the performance of the application is monitored. Then, the performance of that version of the application is compared with the performance of the previous version, and analyzed if there are significant differences.

One problem of the process above is that the performance test must be executed several times to reduce variation in the monitored data. Especially in the DevOps world, where the deployment rate of applications is often considerably higher than in the ``traditional'' software world, the repeated execution of performance tests for each application update poses severe operational challenges.

In my presentation, I will give an overview of the approaches that we study to overcome these challenges and make PRA feasible in the DevOps setting.


\abstracttitle{Exploiting with Integrity---Mining User Data to Improve Software Engineering in the Light of Information Ethics}
\abstractauthor[Markus Borg]{Markus Borg (RISE SICS AB, Lund, Sweden, Markus.Borg@ri.se)}

\license

Software-intensive products developed in market-driven contexts must quickly respond to requirements from users. Also, delivering software that meets the users' expectations of quality is fundamental. Contemporary approaches to tackle the ``need for speed'' include agile and lean software development \cite{DBLP:journals/infsof/DingsoyrL16}, and more recently development with a strong focus on data from operations has gained attention through DevOps \cite{DBLP:conf/hicss/LwakatareKSKOBO16}. There is often a considerable gap between development and operations, but bridging the two activities has the potential to support both responsiveness to market expectations and software evolution in general.

Previously, we have directed considerable research effort toward closing another gap in software engineering, namely between requirements engineering (RE) and testing \cite{DBLP:journals/ese/BjarnasonRBUERSLGF14}. Aligning RE and testing, the ``two ends'' of traditional software development projects, is critical to ensure efficient development of high-quality software---such alignment is important also in iterative development contexts. Among other solution proposals, we have explored supporting RE and testing alignment by a data-driven approach, i.e., identifying connections between resolved issue reports and requirements and test cases \cite{BorgPhd}. By applying machine learning in the footprints of previous issue resolution activities, we provided recommendations of which requirements would be affected during software evolution, and which test cases should be executed to verify the changes.

Now we turn our attention away from issue reports and instead target user data from operations. Our new research direction is still related to recommendation systems highlighting requirements and test cases. However, instead of supporting changes during issue resolution, we plan to predict user needs. In particular, we aim at supporting: i) RE by prioritizing features, and ii) regression testing by improving test case selection. Our approach to tackle this challenge is to continuously mine usage and project data, i.e., data external and internal to the projects. We aim at developing a demonstrator using a mobile platform made available by an industry partner --- a company already specializing in distributed data collection. The envisioned data collection includes: position data, data transfer, CPU workload, and logged user actions.

While there will be technical challenges involved in the research, such as generating actionable recommendations from large amounts of user data, we believe the challenges could be overcome during the project. We should be able to develop a demonstrator that continually feeds data mined from operations directly to a recommendation system for RE and testing activities. On the other hand, another important consideration surfaces: should we do it? Mining rich user data threatens the privacy of the users, which over time could undermine the users' trust. How much privacy are users willing to give up to improve RE and testing? How does it vary depending on how the data is going to be used? To properly address these issues, we collaborate with socio-legal researchers specializing in studies on trust in the digital society \cite{DeKaminski2013}. By combining technical and socio-legal aspects, we hope to bring valuable contributions to DevOps research.

\abstracttitle{Can We Make Performance Visible to Developers?}
\abstractauthor[Lubomir Bulej]{Lubomir Bulej (Charles University Prague, Czech Republic, lubomir.bulej@d3s.mff.cuni.cz)}

\license

The phrase ``Premature optimization is the root of all evil.'' has become a common wisdom and a best practice in software development. Randal Hyde even argues \cite{Hyde:2009:FPO:1569886.1513451} that the phrase has become an excuse for not caring about performance at all, under the assumption that performance is a mere optimization, and that we can always go back and optimize the problematic code.

In his 1974 article for ACM Computing Surveys \cite{DBLP:journals/csur/Knuth74}, Donald Knuth actually wrote: ``We should forget about small efficiencies, say 97 \% of the time: premature optimization is the root of all evil. Yet we should not pass up our opportunities in that critical 3 \%.'' In the same paper, Knuth also wrote: ``In established engineering disciplines, 12 \% performance improvement, easily obtained, is never considered marginal, so why would so many people in computer science pronounce it insignificant?'' Apparently, good engineering practice is to understand the system performance and not give it up through sloppiness. Yet in computer science, or software development in general, we have been often doing exactly that---because of the Moore's law, spending human time to make programs fast was more expensive than buying faster hardware every other year.

However, the scale and complexity of software systems has increased to a point where performance is not a local issue anymore. If we consider the critical 3 \% that Knuth mentioned in his paper---in contemporary systems, those 3 \% may be a lot of code spread throughout the whole software stack that no single person can completely understand anymore. Performance has become deeply ingrained in the software architecture and design, and fixing a performance problem often requires making design changes at different levels of abstraction.

One problem with performance is that is largely invisible to developers, and what is invisible is difficult to manage. In the past, this was also true for software quality, but it has changed when testing became a best practice in modern software development. Testing is not a silver bullet, but it makes certain aspects of quality immediately visible to developers (who dislike failing tests). Moreover, incorporating testing into software development exerts pressure on making code testable, which improves low-level software design. It also makes refactoring safe (rather than feared), which allows developers to evolve software design to match the actual requirements. In the end---besides the testing itself which tells us that a certain feature works as expected---software testing has improved software quality in many other ways by making quality visible.

Dealing with software performance is difficult. While performance can be (with certain difficulties) measured reasonably well (a necessary condition to making it visible and manageable), it is also difficult to interpret. Performance measurements cannot be easily interpreted in a binary fashion and the measurements we reason about have to correspond to a workload that is relevant. While we can think of ways to construct performance tests (similar, but not quite like functional unit tests) that can be evaluated automatically, it is the relevance of these tests that matters, which is where the DevOps culture could help. If the Devs can find (through Ops) what is relevant to performance, then they can write relevant performance tests, and thus make performance visible in the software development (and possibly continuous deployment) process. Can we help make performance visible?



\abstracttitle{Dealing with Uncertainty in Developer Targeted Analytics}
\abstractauthor[J\"urgen Cito]{J\"urgen Cito (University of Zurich, Switzerland, cito@ifi.uzh.ch)}

\license


Runtime information of deployed software has been used by business and operations units to make informed decisions under the term  ``analytics''. However, decisions made by software engineers in the course of evolving software have, for the most part, been based on personal belief and gut-feeling \cite{DBLP:conf/sigsoft/CitoLFG15}. This could be attributed to software development being, for the longest time, viewed as an activity that is detached from the notion of operating software in a production environment. In recent years, this view has been challenged by the emergence of the DevOps movement, which aims to promote cross-functional capabilities of development and operations activities within teams. This shift in mindset requires analytics tools that specifically target software developers. We investigate how to support developers in their decision-making process by incorporating runtime information in source code (``Developer Targeted Analytics'') \cite{DBLP:conf/oopsla/CitoLGDKR15}. In this approach, we also provide live feedback in IDEs by inferring the impact of code changes on software performance.
However, no prediction is perfect. Every inference model comes with a level of uncertainty. To make integrated runtime feedback a proper basis for decision-making, we need methods to properly quantify uncertainty and consequently communicate it to software developers. In this seminar, I want to discuss sources of variation in software performance (e.g., cloud instance variability \cite{DBLP:journals/toit/LeitnerC16}) and possible ways to model uncertainty. Further, I want to review different ways to communicate runtime feedback through visualizations.



\abstracttitle{Transforming Operations for the Cloud}
\abstractauthor[Georgiana Copil]{Georgiana Copil (TU Vienna, Austria, e.copil@dsg.tuwien.ac.at)}

\license

In a culture where development and operations merge, sustained by new computing models such as cloud computing and management automation technologies, the operations processes need to evolve. Although extensive work is performed addressing the perspective of the cloud provider \cite{IBM2010ISM}, IT service management, and particularly operation management, is disregarded for the cloud customer side.

Well established standards like ITIL \cite{ArrajITILTheBasics}, BSI ISO 20000 \cite{ISO20000}, and FitSM \cite{FitSM} are currently used for IT service management. However, these standards need to be adapted or completely re-designed to accommodate disruptive trends in technology, as emphasized by Forsgren et al. \cite{ForsgrenHumble2016DevOpsITSM} and Fuggetta et al. \cite{DBLP:conf/icse/FuggettaN14}. A plethora of configuration management and automation tools (e.g., Chef, Puppet, Ansible, Brooklin, Vagrant, or rSYBL \cite{DBLP:conf/icsoc/CopilMTD13}) are being created for providing support both in development and operation. In companies using cloud services for developing their products, some organization roles will have to change, or be replaced by automated tools. For instance, procurement engineers might need to work with configuration management engineers in order to keep track with the costs that depend on system's configuration, or might simply be discarded from the organization chart.
Although change is needed in processes and service management organizational charts, tools should be adapted as well in order to integrate better within the organization. For instance, controllers could notify the responsible operation engineers or managers \cite{DBLP:conf/icsoc/CopilTD15}, with the correct content and frequency, on abnormal behavior, or on detected incidents. VRealize  and AppDynamics  provide role-based alerting, and so-called ``Virtual War Rooms'' for enhancing Dev \& Ops collaboration. Although these are a promising start, further investigation is necessary for understanding how operation-time events of various types (e.g., changes, faults, or incidents) can be automatically classified, and used for predicting such events, and analyzing root causes, and providing tighter connection with the developers or operators involved in these aspects.

The next challenging aspect is the decoupled ownership of operations management, and especially of incident management, due to the decoupled ownership of cloud services, and systems using these services \cite{DBLP:conf/srds/ArefinJ11}. As stated before, systems using cloud services would need an adapted IT Service management processes, and integrate information coming from cloud providers into these processes. However, the reduced control over data \cite{zawoad2012towards,DBLP:journals/tdsc/ZawoadDH16}, absence of standard format and logs \cite{DBLP:journals/tdsc/ZawoadDH16}, multi-tenancy \cite{zawoad2012towards}, make this integration very challenging. Better operation management possibilities on the cloud customer side, which includes information related to cloud services operation, can reduce the trust concerns and increase the adoption of cloud technologies.

\abstracttitle{Challenges in Architectural Modeling for Performance-aware DevOps}
\abstractauthor[Robert Heinrich]{Robert Heinrich (Karlsruhe Institute of Technology, Germany, robert.heinrich@kit.edu)}

\license

Building software applications by composing cloud services promises many benefits such as flexibility and scalability. However, it leads to major challenges like increased complexity, fragility and changes during operations that cannot be foreseen in development phase. Cloud-based applications change rapidly and thus require increased communication and collaboration between software developers and operators as well as strong integration of building, evolving and adaptation activities.

While previous research focused on automated system adaptation, increased complexity, heterogeneity and limited observability, makes evident that we need to allow operators (humans) to engage in the adaptation process. Architectural models such as those of the Palladio approach \cite{reussner2016modeling} are a foundation for involving humans and conducting analysis, e.g., for performance and privacy. During operations the system often drifts away from its development models. Run-time models are kept in-sync with the underlying system. However, typical run-time models are close to an implementation level of abstraction which impedes understandability for humans.

DevOps practices enable software developers and operators to work more closely \cite{DBLP:books/daglib/0036722}. The software application architecture is a central artifact for developers and operators. New architectural styles such as microservices are proclaimed to satisfy requirements like scalability, deployability and continuous delivery. By merely introducing new architectural styles, however, current problems in the collaboration and communication among stakeholders of the development and operations phases are not solved. Existing architectural models used in the development phase differ from those used in the operation phase in terms of purpose (finding appropriate design vs. reflecting current system configurations), abstraction (component-based vs. close to implementation level) and content (static vs. dynamic). These differences result in limited reuse of development models during operations and limited phase-spanning consideration of the software architecture.

We are developing the iObserve approach to address architectural challenges in \linebreak performance-aware DevOps. iObserve provides a megamodel \cite{DBLP:journals/sigmetrics/Heinrich16} to bridge the divergent levels of abstraction in architectural models used in development and operations. We employ descriptive and prescriptive architectural run-time models for realizing the MAPE loop. Including dynamic content, like in-memory objects and their communications, will help operators to observe and adapt the application when anomalies exceed automated planning routines \cite{heinrich2017a}.



\abstracttitle{Efficient Resilience Benchmarking of Microservice Architectures}
\abstractauthor[Andr\'{e} van Hoorn]{Andr\'{e} van Hoorn (University of Stuttgart, Germany, van.hoorn@informatik.uni-stuttgart.de)}

\license

The microservice architectural style \cite{Newman2015BuildingMicroservices} is gaining more and more prevalence in industrial practice when constructing complex, distributed systems. One of its guiding principles is design for failure, which means that a microservice is able to cope with failures of other microservices and its surrounding software/hardware infrastructure. This is achieved by employing architectural patterns such as circuit breaker and bulkhead \cite{Nygard2007ReleaseItDesignAndDeployProductionReadySoftware}. 

Resilience benchmarking \cite{DBLP:books/sp/wolter2012/VieiraMSK12} aims to assess failure tolerance mechanisms---for instance, via fault injection \cite{Natella:2016:ADS:2856149.2841425}. Meanwhile, resilience benchmarking is not only conducted in development and staging environments, but also during a system's production use \cite{Krishnan2012WeatheringTheUnexpected}---for instance, via Netflix's Simian Army \cite{Netflix2016SimianArmy}. Existing resilience benchmarks for microservice architectures are ad-hoc and based on randomly injected faults.

In this talk, I will sketch the vision for efficient resilience benchmarking of microservice architectures. Resilience vulnerabilities shall be detected more efficiently, i.e., faster and with fewer resources, by incorporating architectural knowledge as well as knowledge about the relationship between performance/capacity/stability (anti) patterns and suitable injections. The idea builds on existing works on model-based and measurement-based dependability evaluation of component-based software systems.



\abstracttitle{Benchmarking Quality of Performance Evaluation in the DevOps World}
\abstractauthor[Vojt\v{e}ch Hork\'{y}]{Vojt\v{e}ch Hork\'{y} (Charles University Prague, Czech Republic, vojtech.horky@d3s.mff.cuni.cz)}

\license

Performance in the DevOps world today typically revolves around the issue of collecting performance data from running applications and analyzing them (by the dev-and-ops) to detect performance anomalies. But are we able to say what is the ratio of detected and undetected issues, i.e., how reliable are the tools we use?

The nature of continuous delivery complicates answering this. The software is updated too often---we are not able to collect enough information before a new version is introduced and we do not care that much about regressions in a week-old version.
So how about writing a benchmark emulating the production of a DevOps-driven application? The benchmark would allow us to evaluate quality of tools and approaches we use.
\begin{itemize}
\item The benchmark provides the baseline truth and we can quantify how many issues went undetected. Ranking the issues (e.g., how critical a problem is) gives even more information about the precision of the tested approach.
\item We can (at least partially) evaluate the solution without touching the production environment (the agility of DevOps suggests that we try the tools directly but that may not be always possible).
\item Having a benchmark would allow a comparison of different approaches that is otherwise virtually impossible because of the variances between the systems under test.
\end{itemize}
The benchmark can test the solution from three different angles. As a database of measurements of various software components in different versions it evaluates how well are we able to detect anomalies. Also the benchmark could emulate a volatile environment where new versions of different components are loaded over time---a score indicates whether the measurement framework survives in such environment and how precise results it provides. Model-based approaches can be used here as well. And finally such emulation could also be used for testing the continuous delivery infrastructure and related processes---effectively benchmarking quality assurance processes.

The current status is somewhere at the level of ``collecting ideas for implementation''. The emulated application shall be component-based with a well-defined architecture; a good start might be the CoCoME~\cite{DBLP:conf/dagstuhl/2007cocome} or PNMR~\cite{DBLP:conf/wosp/BrunnertDK15}. We would then need to create several versions differing at various levels and add some kind of self-measurement infrastructure. Then we can quantify how well the tested solution adapts to the changes and how precise data it produces~\cite{DBLP:conf/wosp/HorkyKLT16}.
Collecting data for the database of measurements is the easier step---the difficult one is establishing the base truth: what were the actual regressions. Around the database we would then create a harness where individual solutions could be plugged-in and run on different data sets.
Looking forward to see whether this might be interesting for someone else as well.



\abstracttitle{Machine Learning Meets DevOps}
\abstractauthor[Pooyan Jamshidi]{Pooyan Jamshidi (Imperial College London, United Kingdom, p.jamshidi@imperial.ac.uk)}

\license



Today's mandate for faster business innovation, faster response to changes in the market, and faster development of new products demand a new paradigm for software development. DevOps is a set of practices that aim to decrease the time between changing a system in Development, and transferring the change to the Operation environment, and exploiting the Operation data back in the Development \cite{DBLP:books/daglib/0036722}. DevOps practices are typically relying on large amount of data coming from Operation. The amount of data depends on the architectural style \cite{DBLP:journals/software/BalalaieHJ16}, the underlying development technologies and deployment infrastructure. For instance, big data distributed systems consist of an extensible execution engine (e.g., MapReduce), pluggable distributed storage engines (e.g., Apache Cassandra), and a range of data sources (e.g., Apache Kafka). Each of these produce a considerable amount of data regularly in each fraction of second.

However, in order to make effective decisions in Development, e.g., architectural refactoring in order to make the system architecture sustainable \cite{pahl2017cloud} during its lifetime, there has to be efficient processing of such large amount of data in place in order to process the operational data. In this situation where data streams are increasingly large-scale, dynamical and heterogeneous, mathematical and algorithmic creativity are required to bring statistical methodology to bear. Statistical machine learning merges statistics with the computational sciences. Statistical machine learning can fill the gap between operation and development with some more efficient analytical techniques. The data efficient techniques can provide more deep knowledge and can uncover the underlying patterns in the operational data in order to detect anomalies in the operation or detect performance anti-patterns \cite{bersani2016continuous}. This knowledge can be very practical if detected ontime in order to refactor the development artifacts including code, architecture and deployment \cite{XX-SPEC-RG-2015-001-DevOpsPerformanceResearchAgenda}.

In this talk, I present our recent work on configuration tuning of big data software, where I primarily applied Bayesian Optimization and Gaussian Processes, a data efficient statistical machine learning method, in order to quickly find optimum configurations \cite{DBLP:journals/corr/JamshidiC16}. I also talk about transfer learning to exploit complimentary and cheap information (e.g., past measurements regarding early version of the system) to enable learning accurate models quickly and with considerably less cost~\cite{JVKSK:SEAMS17,JSVKPA:ASE17}. 


\abstracttitle{Evaluating the Effectiveness of Different Load Testing Analysis Techniques}
\abstractauthor[Zhen Ming (Jack) Jiang]{Zhen Ming (Jack) Jiang (York University, Toronto, Canada, zmjiang@cse.yorku.ca)}

\license


Large-scale software systems like Amazon and eBay must be load tested to ensure they can handle hundreds and millions of current requests in the field. Load testing usually lasts for a few hours or even days and generates large volumes of system behavior data (execution logs and counters). This data must be properly analyzed to check whether there are any performance problems in a load test. However, the sheer size of the data prevents effective manual analysis. In addition, unlike functional tests, there is usually no test oracle associated with a load test. To cope with these challenges, there have been many analysis techniques proposed to automatically detect problems in a load test by comparing the behavior of the current test against previous test(s). Unfortunately, none of these techniques compare their performance against each other.

In this talk, I describe our work on the empirical evaluation of the effectiveness of different test analysis techniques \cite{DBLP:conf/icst/GaoJBL16}. We have evaluated a total of 23 test analysis techniques using load testing data from three open source systems. Based on our experiments, we have found that all the test analysis techniques can effectively build performance models using data from both buggy or non-buggy tests and flag the performance deviations between them. It is more cost-effective to compare the current test against two recent previous test(s), while using testing data collected under longer sampling intervals ($\geq$ 180 seconds). Among all the test analysis techniques, Control Chart, Descriptive Statistics and Regression Tree yield the best performance. Our evaluation framework and findings can be very useful for load testing practitioners and researchers. To encourage further research on this topic, we have made our testing data publicity available to download.



\abstracttitle{How I Learned to Stop Worrying and Love Capacity Shortages}
\abstractauthor[Cristian Klein]{Cristian Klein (Umea University, Sweden, cklein@cs.umu.se)}

\license

My research focuses on engineering cloud applications so as to maintain responsiveness despite infrastructure capacity shortages. This has several benefits. First, the risk of user experience degradation is reduced when the application observes a sudden increase in popularity. Second, it allows the infrastructure to operate at higher utilization, which helps reduce costs. The main idea is to mark certain code of the application as optional and selectively deactivate its execution, as required to maintain a target response time. The problem can be decomposed into two questions: What code to deactivate and when to deactivate such code?

To answer the ``what'' question, we proposed a methodology to retrofit admission control into existing cloud applications~\cite{Chaudhry970648}. The first step is to model the user behavior based on production logs in a manner that is both realistic, but also flexible enough to test various ``what-if'' scenarios. Next, the model can be used to generate an amplified workload, identify bottlenecks and add circuit breakers around such bottlenecks, as directed by business objectives. Several iterations can be performed to add multiple circuit breakers, until the resilience of the application is judged to be sufficient. We evaluated the methodology against a production cloud application, featuring a large code base.
 
To answer the ``when'' question, we proposed ``brownout'' a software engineering methodology to decide when to disable optional content, such as recommendations or comments. We used control-theory to design a controller that measures application response time and adjusts the probability of serving optional content, as required to maintain a given target response time \cite{DBLP:conf/icse/KleinMAH14,Desmeurs:2015:EAB:2861850.2861966}. While brownout itself is rather uninstrusive to the developer, the application's struggle with capacity shortage can no longer be observed by measuring utilization or response time, hence brownout-unaware components around the application may take errornous decisions. To tackle this issue, we proposed a brownout-aware load-balancer~\cite{7024056} and admission controller~\cite{DBLP:conf/srds/KleinPDDMAHE14}.

Our techniques allow software engineers to deprioritize performance-related non-functional requirements, such as scalability, and reduce time-to-market. This is important in the context of lean thinking, which advocates the creation of minimum viable products to discover customer requirements.

\abstracttitle{Performance Modeling Challenges while Modernizing Existing Software towards Microservices}
\abstractauthor[Holger Knoche]{Holger Knoche (Christian-Albrechts-Universit\"at zu Kiel, Germany, hkn@informatik.uni-kiel.de)}

\license

In order to fully realize the promises of DevOps, namely the fast and continuous delivery of high-quality software, an appropriate software architecture is required. Currently, Microservices are considered the premier software architecture for this purpose \cite{DBLP:books/daglib/0036722}. As a consequence, many companies consider the introduction of Microservices to their existing software by implementing new features as Microservices or replacing existing functionality by Microservices. Further information on this topic can be found in \cite{DBLP:journals/software/BalalaieHJ16,Newman2015BuildingMicroservices}.
Due to their highly distributed nature, Microservices introduce several new challenges for both development and operations. Developers now have to cope with potential performance degradation due to remote invocations, the possibility of partial failure, and the loss of ACID transactionality \cite{DBLP:conf/wosp/Knoche16}.  From an operations point of view, deploying, running, and monitoring a large number of service instances pose the greatest challenge.
To address the development challenges described above, performance simulations are a promising option. As the mentioned effects are largely determined by the service design, the use of design-time models such as the Palladio Component Model \cite{DBLP:journals/jss/BeckerKR09} presents itself. The operational challenge is addressed by a group of tools which facilitate the deployment and operation of highly distributed applications, the most prominent being Kubernetes \cite{Kubernetes} and Mesos \cite{DBLP:conf/nsdi/HindmanKZGJKSS10}. These tools automatically deploy services to a pool of nodes based on given constraints, and dynamically adjust the deployment should it become necessary (e.g., due to node failure).
Unfortunately, however, current design-time performance models lack an appropriate abstraction for such dynamic, self-adaptive deployments. What adds further difficulty is the fact that in migration contexts, both ``traditional'' applications and Microservices interact with each other, and the performance impact of this interaction is of particular importance. Therefore, a formalism capable of modeling and simulating both worlds in an appropriate way is required.
In my talk, I am going to further illustrate the performance challenges of migrating towards Microservices using examples from industrial practice. Furthermore, I will give a brief presentation of Mesos as an example of a modern deployment platform, and point out the performance modeling challenges that arise from such platforms.



\abstracttitle{The Importance of Data Science for DevOps and Continuous Delivery}
\abstractauthor[Philipp Leitner]{Philipp Leitner (University of Zurich, Switzerland, leitner@ifi.uzh.ch)}

\license

Academic research in software engineering (SE) is at a pivotal point. With the advent of DevOps, the SE research community needs to expand and augment its traditional topics (e.g., requirements elicitation, software architecture, formal specification, or testing), to include new runtime-related challenges. These include, but are not limited to, cloud computing, continuous delivery and deployment (CD), edge computing and the Internet of Things, or the Facebook-style ``move fast'' development philosophy that emphasizes the importance of quick delivery over strenuously verified correctness.

One implication of this new ``runtime focus'' is that SE can, and will, become more data-driven. Today, however, data science in SE is still the domain of dedicated specialists more than part of day-to-day development. Developers still struggle to make systematic use of the deluge of data available through Application Performance Monitoring tools \cite{DBLP:conf/sigsoft/CitoLFG15}, rollout decisions in continuous deployment and live testing are based on intuition rather than collected empirical evidence \cite{DBLP:journals/peerjpre/SchermannCLZG16}, and software performance engineering techniques are still not widely found in industrial practice \cite{leitner:17}. We argue that one reason for this is that performance monitoring and analysis tools are currently not ``developer-targeted''. Their results and visualizations are not actionable for developers, and they do not provide robust value without expert knowledge in statistics and data science.

In my talk, I will focus on three DevOps-related SE topics and how they relate to data science. Firstly, I will (briefly) introduce our work on Feedback-Driven Development (FDD), a concept that aims to make runtime performance data more useful for software developers \cite{DBLP:journals/peerjpre/SchermannCLZG16}. Secondly, I will discuss the importance and challenges of data science for Continuous Delivery, Continuous Deployment, and live testing. In this segment, I will focus on the trade-off between release velocity and confidence \cite{DBLP:journals/peerjpre/SchermannCLZG16}. Finally, I will discuss the challenge of cost-aware operation of applications on top of public IaaS clouds. I will present recent results in this domain \cite{DBLP:journals/toit/LeitnerC16}, and discuss problems and challenges.

\abstracttitle{Industrial-grade DevOps: DevOps in the Digitalized Industrial World}
\abstractauthor[Fei Li]{Fei Li (Siemens AG, Austria, lifei@siemens.com)}

\license

Industrial systems, such as factory automation, infrastructure management, utility management, require highly reliable and near real-time solutions. The lifespan of such systems is long, and traditionally, during their lifetime the focus of software development and operation is to ensure high reliability and performance, and to continuously comply with various regulations. Therefore, new versions are released on a yearly basis or even less frequently, and whole-sale update has to be carried out for each release. Correspondingly, monolithic architecture is dominant among industrial software.

However, the recent digitalization movement in industrial systems presents significant challenges to the traditional way of industrial software delivery. In a digitalized industrial world, flexibility and time to market are the key to ensure business success, while at the same time the requirements on software quality cannot be compromised. The current development methodology, software architecture and tools cannot support such demands in a digitalized industrial world.

To this end, the concept of Industrial-grade DevOps has been proposed by the Corporate Technology department of Siemens to address the challenges of software delivery in the digitalized world. The vision of industrial-grade DevOps is to ``move from monolithic products with long release cycles and wholesale updates to componentized products with continuous feature release and evolution in run.''

In this scope, the following key research topics will be investigated.

\begin{enumerate}
\item Microservice architecture in industrial systems. The key is to ensure software quality while continuously evolving the microservice architecture. In particular, we focus on five software qualities:
\begin{itemize}
\item Availability---Throughout continuous updates of independent microservices
\item Roll-backability---To roll back software and physical systems to earlier states in case of update failure
\item Resilience---In case single or even multiple microservices fail, the system as a whole should not fail
\item Security---To secure information as well as physical assets
\item Changeability---To understand the impact of individual changes of independent changes
\end{itemize}

\item Knowledge-drive architectural decision making methods. Development and operational knowledge is the cornerstone to make optimal decisions in the fast evolving DevOps environments.
\begin{itemize}
\item Extracting operational routines, workflows and data structures as patterns and document them like architectural patterns
\item Building knowledge base that associates development and operational activities with software qualities.
\item Use operational patterns to address requirements from Ops
\item React to quality problems by using the result of operational data analysis
\end{itemize}

\end{enumerate}


\abstracttitle{The Challenges and Benefits of Synthesizing and Theorizing the DevOps Phenomenon in Software Engineering}
\abstractauthor[Lucy Ellen Lwakatare]{Lucy Ellen Lwakatare (University of Oulu, Finland, Lucy.Lwakatare@oulu.fi)}

\license

Software-intensive companies constantly try to improve their software development process for better software quality and a faster time to market. The continuous delivery paradigm represents a more recent mainstream software development practice that is predominant in the web domain and has a huge potential in its adoption in other domains \cite{DBLP:conf/icsob/KarvonenLSBOKO15,DBLP:conf/hicss/LwakatareKSKOBO16}.

DevOps---collaboration between development and operations activities---is necessary for adopting and enabling continuous delivery and deployment. However, the state-of-art on the DevOps phenomenon is driven by industry, and has very limited contributions in research \cite{DBLP:conf/xpu/LwakatareKO15}.  This is evident in the article on a systematic mapping study of continuous deployment that has also shown that the research is lagging behind in systematizing knowledge and in validating many of the claims advocated in continuous deployment and DevOp practices \cite{Rodriguez2017263}.

Using a multi-vocal ``grey'' literature review and case study approaches in software engineering, a more focused study on the DevOps phenomenon is conducted with the aim of synthesizing, systematizing and providing evidence of DevOps practices \cite{lwakatare2016exploratory}. The findings show that, even though it is possible to abstract the wide and diverse set of practices advocated by DevOps into useful patterns that can be adopted by other companies seeking to implement DevOps, from the research respective it remains challenging to generate theories.

\abstracttitle{Releasing High-performance Software More Rapidly with Lower Costs using Continuous Test Optimization}
\abstractauthor[Dusica Marijan]{Dusica Marijan (Simula, Norway, dusica@simula.no)}

\license

Organizations that follow DevOps practice often benefit from improved efficiency, more reliable releases, higher product quality, or better user experience. However, successfully implementing DevOps entails a number of challenges. One such challenge includes implementing an automated and cost-effective continuous testing, as a prime driver for DevOps. We analyze this challenge in the context of testing industrial communication systems. In this context, any supporting technique/tool has to integrate within CI/CD workflows, enable automation, and support collaboration between development and QA. In our work, we are interested in techniques for reaching a high degree of test automation and optimization for a given production environment, to enable faster deployments and more frequent releases of features and bug fixes, ultimately enabling rapid releases of products. We analyze how can historical data, usage profiles, sampling-based techniques, and other techniques within and outside of a production environment be efficiently used to collect performance metrics, recognize patterns and performance bottlenecks, in order to parameterize optimization algorithms for improved and more cost-effective testing.


\abstracttitle{Joining Adaptation and Evolution Control Loops to Manage Performance in a DevOps Setting}
\abstractauthor[Claus Pahl]{Claus Pahl (Free University of Bozen-Bolzano, Italy, claus.pahl@unibz.it)}

\license

An important problem is how to organise and manage the different performance engineering tasks within a DevOps setting. The core of the proposal here is a process model combining an adaptation and evolution loop of systems change, which is an extension of the ADEPS model \cite{DBLP:conf/ecsa/WeynsCVK15} that joins two feedback loops over time. It consists of two intertwined change spirals, drawn out over time:
An evolution helix reflecting long-term changes, usually with clear reference back into a development stage
An adaptation helix reflecting short-term changes, usually dynamically in systems in an automated fashion.
This change process model is governed by goal continuity as the key principle: in dynamic systems, performance goals need to be validated and maintained continuously. However, goal continuity is challenged by uncertainties arising from stakeholders, platform and information incompleteness and inaccuracy. Goal continuity is enabled through feedback that links Operations back into Development concerns if required. Tools and mechanisms that enact this are 
(i) controllers of dynamic adaptation in the Operations domain (e.g., through platform adaptation) and
(ii) experiments and prototyping in continuous evolution in the Development domain (e.g., as part of refactoring and re-architecting activities)

Both are critical for the enablement of performance engineering. In some way, a MAPE-K loop is in place for both feedback loops, one in an autonomous setting, the other in a more human controlled form. What is needed is a uniform model framework to manage both.

Our objective should be to share the same mechanisms of monitoring, analysis and decision making consistently for both cycles. What a controller does for autonomic computing needs to be provided by managed experiments, prototypes and recommenders for an evolutionary scenario. A joint formal, methodological and technical basis shall be the aim to support both with the same rigour. We found for instance an abstraction of problem situations in terms of patterns to be beneficial \cite{DBLP:conf/icsoc/JamshidiPCL14,DBLP:conf/qosa/JamshidiSPAME16}. The proposal is therefore to identify a set of common performance management models, which can build on these patterns.
The wider context of this model, i.e., what drives the DevOps process, also needs to be understood better. Technical and economic sustainability is the aim of Continuous Development and Operations \cite{DBLP:conf/icse/FitzgeraldS14}. A software system is sustainable if it is resilient to emerging uncertainty. Cost and performance are intertwined.

Cloud computing shall be selected as a specific case that highlights the relevance of performance engineering in a service-oriented architecture that is constrained by the service-level agreements between solution service provider and consumer.



\abstracttitle{Performance Engineering in Fog Computing --- An Overview}
\abstractauthor[Stefan Schulte]{Stefan Schulte (TU Wien, Austria, s.schulte@infosys.tuwien.ac.at)}

\license


First coined by Cisco \cite{DBLP:conf/sigcomm/BonomiMZA12}, the term fog computing describes the usage of well-known principles from cloud computing at the edge of the network, most importantly the virtualization of resources provided by networked devices. By applying fog computing, it is possible to lease and release networked devices as virtualized assets in an on-demand, utility-like fashion and to enable rapid elasticity through scaling these leased assets up and down, if necessary. Also, fog devices may be orchestrated in order to cooperatively process data~\cite{DBLP:conf/soca/Skarlat0BL16}. Fog computing partially resembles ideas of mobile computing, however it does focuses on Internet of Things (IoT) entities like sensor nodes, smart objects, or cyber-physical systems. Apart from the term fog computing, edge computing is also frequently used to describe this approach.

One main motivation for the advent of fog computing is its alignment with the basic structure of the IoT: Within the IoT, a plethora of technologically heterogeneous, connected devices are interacting with each other \cite{DBLP:journals/cn/AtzoriIM10}. While IoT devices are highly (geo-)distributed, cloud resources are actually provided in quite a centralized way. This is not surprising, since the success of cloud computing can be attributed (amongst other reasons) to the economies of scale achieved in centralized, very large data centers. In contrast, fog resources are located at or nearby the edge of the network and thus in the vicinity of IoT devices. By exploiting already available resources, it is possible to do computational tasks ``on-site'', i.e., close to the data sources and/or sinks, instead of executing these tasks in the cloud. For instance, preprocessing in data stream scenarios or data prefiltering in Big Data scenarios may be done in the fog instead of the cloud. Actually, decreased latency is one major reason for the usage of fog-based computational resources~\cite{DBLP:conf/sigcomm/BonomiMZA12}, therefore fog computing is both an enabler and beneficiary of performance engineering.

Since fog computing is still a very recent research topic, there are (to the best of our knowledge) no dedicated approaches for performance engineering in the fog. In fact, there are not even DevOps approaches for fog-based applications yet. This talk will therefore present open questions in the field, discussing also the differences between fog and cloud computing and why there is a need for specific performance engineering in the fog. In addition, some first ideas towards solving these questions will be presented.

For this, we will introduce the notion of glocal living applications (GLAs), which are fog- and cloud-native applications by design. By applying container technologies, GLAs are isolated, portable applications, which can be hosted on any virtualized infrastructure. GLAs are able to move freely within and between data centers, both in the cloud and in the fog, e.g., by acquiring computing resources from a different cloud service or cloud region closer to the customer, or by offloading computational tasks from the fog to the cloud and vice versa, through migrating application code via the mobile code paradigm. This allows GLAs to minimize latency and enables region-specific service customization.



\abstracttitle{Improving the Performance of Database-centric Applications through DevOps}
\abstractauthor[Weiyi Shang]{Weiyi Shang (Concordia University, Canada, shang@encs.concordia.ca)}

\license

There is a growing gap between the software development and operation, especially for database-centric applications. Software developers of database-centric applications typically leverage Object-Relational Mapping frameworks such as Hibernate to ease the access of database, and caching frameworks to optimize the performance of database access. However, the use of such framework brings extra challenges, such as performance overhead. Developers may not understand the field performance impact of executing the source code due to the use of the complex Object-Relational Mapping frameworks. On the other hand, developers often configure the caching framework based on their own experiences and gut feelings. A suboptimal configuration can cause even worse performance than without leveraging any caching framework.

The introduction of DevOps bridges the gap between software development and operation and brings developers the rich field information, such as actual impact of source code execution in the field and the usage of the system by real users. With the field information, we propose techniques that can improve the performance of database-centric applications. In particular, one technique leverages the field data from end users to optimize the configuration of caching frameworks~\cite{DBLP:conf/sigsoft/ChenSHNF16}. Instead of configuring based on developers' experiences, by leveraging the field information from software operation, developers can optimize the caching framework configuration by knowing the data-access patterns of database tables. Significant performance improvements are shown by evaluating the above technique with open source software and with our industrial partners in practice.



\abstracttitle{SPE Meets DevOps: Best Friends or Consensual Enemies?}
\abstractauthor[Catia Trubiani]{Catia Trubiani (Gran Sasso Science Institute, Italy, catia.trubiani@gssi.infn.it)}

\license


DevOps is a novel trend that aims at bridging the gap between development and operations, while providing the control of deployment and application runtime in the hands of developers. When applied in the context of software performance engineering, it raises new challenges related to which performance data should be carried back-and-forth between runtime and design-time and which feedback should be provided to developers to support them in the diagnosis of performance results. There is an obvious trade-off in the performance evaluation of early model abstractions where detected problems are cheaper to fix but the amount of information is limited, and late performance monitoring on running artifacts, where the results are more accurate but several constraints have been added on the structural, behavioral, and deployment aspects of a software system. The goal of this talk is to point out the research challenges in this domain thus to understand at which extent the meeting between SPE and DevOps leads to make them ``best friends'' by exploiting some synergy or ``consensual enemies'' by discovering some incompatibility.


\abstracttitle{Performance-aware DevOps Through Declarative Performance Engineering}
\abstractauthor[J\"urgen Walter]{J\"urgen Walter (Julius Maximilians Universit\"at W\"urzburg, Germany, juergen.walter@uni-wuerzburg.de)}

\license


Performance is of particular relevance to software system design, operation, and evolution because it has a major impact on key business indicators. Over the past decades, various methods, techniques, and tools for modeling and evaluating performance properties of software systems have been proposed covering the entire software life cycle \cite{XX-SPEC-RG-2015-001-DevOpsPerformanceResearchAgenda}. However, the application of performance engineering approaches to solve a given user concern is still rather challenging and requires expert knowledge and experience. There are no recipes on how to select, configure, and execute suitable methods, tools, and techniques allowing to address the user concerns.
The application of performance engineering approaches is challenging even in classical slow and heavy-weight processes with long-term release cycles. DevOps automates the process of software delivery and infrastructure changes. This results in short-term build and release cycles. Consequently, there is a more frequent need for performance evaluation. 
Reinforced by faster developments cycles, software and system engineering requires an easy, holistic integration and automation of performance engineering techniques. Declarative Performance Engineering (DPE) \cite{DBLP:conf/wosp/WalterHKOK16} aims to reach this decoupling of the description of the user concerns to be solved (performance questions and goals) from the task of selecting and applying a specific solution approach. The strict separation of ``what'' versus ``how'' enables the development of different techniques and algorithms to automatically select and apply a suitable approach for a given scenario. The goal is to hide complexity from the user by allowing users to express their concerns and goals without requiring any knowledge about performance engineering techniques. Realizing the DPE vision my research includes amongst others
\begin{itemize}
\item performance concern language design
\item reference architecture for automated deduction of concerns
\item tool decision support and capability model
\item automated performance model learning
\end{itemize}



\abstracttitle{Monitoring DevOps Processes and Experimental Process Improvement}
\abstractauthor[Ingo Weber]{Ingo Weber (Data61, CSIRO, Australia, Ingo.Weber@data61.csiro.au)}

\license


For the last 4 years, my team has been working in the DevOps space, and the lack of readily available material on this topic motivated us to write our DevOps book \cite{DBLP:books/daglib/0036722}. In the first part of my talk I summarized the key points from the book.
The second part of my talk focused on two topics of our DevOps research that are relevant for performance engineering, specifically (i) correlation of DevOps process event occurrence with changes in metrics, and (ii) applying DevOps methods to process improvement and measuring outcomes.
 
\paragraph*{Correlating metrics with DevOps process events}
 
The Process-Oriented Dependability (POD) framework has been developed to deal with failures of application operations, since they are one of the main causes of system-wide outages in cloud environments. This particularly applies to DevOps operations, such as backup, redeployment, upgrade, customized scaling, and migration that are exposed to frequent interference from other concurrent operations, configuration changes, and resources failure. However, previous practices failed to provide a reliable assurance of correct execution for these kinds of operations. In this work, we devised an approach to address this problem that adopts a regression-based analysis technique to find the correlation between an operation's activity logs and the operation activity's effect on cloud resources. The correlation model is then used to derive assertion specifications, which can be used for runtime verification of running operations and their impact on resources. The research has been published \cite{DBLP:conf/issre/FarshchiSWG15}.
 
\paragraph*{Experimental process improvement}
 
Software systems that support Business Process Management are in widespread use. They play an important role in facilitating process automation and process improvement. Yet, there is hardly any insight into whether the implementation of a supposedly improved process model leads to an actual improvement in the process. The research problem this work addresses is: how can we determine if a new variant of a process model is an improvement over a previous variant, with respect to relevant measures? To this end, we suggest to build on recent software engineering concepts from the DevOps movement, as well as adopt learnings from performance engineering for measurements and analysis. On this basis, we want to develop novel techniques that provide the infrastructure for assessing in how far a specific business process change leads to an improvement. A vision paper on this topic has been published \cite{DBLP:conf/simpda/WeberM15}.



\abstracttitle{Performance of Continuous Delivery Pipelines}
\abstractauthor[Johannes Wettinger]{Johannes Wettinger (University of Stuttgart, Germany, mail@jowettinger.de)}

\license

During the last 3--4 years, my research activities focused on diverse challenges concerning DevOps and continuous delivery. I was diving into these topics as part of my works and efforts as a PhD candidate at the Institute of Architecture of Application Systems (IAAS) at the University of Stuttgart. The motivation for these research topics is obvious: especially users, customers, and other stakeholders in the fields of Cloud services, Web applications, mobile apps, and the Internet of things expect quick responses to changing demands and occurring issues. Consequently, shortening the time to make new releases available becomes a critical competitive advantage. In addition, tight feedback loops involving users and customers based on continuous delivery ensure to build the ``right'' software, which eventually improves customer satisfaction, shortens time to market, and reduces costs.

Independent of the chosen approach to establish continuous delivery by tackling cultural and organizational issues, a high degree of technical automation is required. This is typically achieved by implementing an automated continuous delivery pipeline (also known as deployment pipeline), covering all required steps such as retrieving code from a repository, building packaged binaries, running tests, and deployment to production. Such an automated and integrated delivery pipeline improves software quality, e.g., by avoiding the deployment of changes that did not pass all tests. Moreover, the high degree of automation typically leads to significant cost reduction because the automated delivery process replaces most of the manual, time-consuming, and error-prone steps. Establishing a continuous delivery pipeline means implementing an individually tailored automation system, which considers the entire delivery process. Furthermore, a separate pipeline has to be established for each independently deployable unit, e.g., a monolith or microservice. As a result, a potentially large and growing number of individual pipelines have to be established. Therefore, my research focuses on dynamically and systematically establishing such continuous delivery pipelines.

The key building blocks of a continuous delivery pipeline are diverse application environments such as a development environment, test environment, and production environment. A specific goal is to provide discovery, consolidation, utilization, and orchestration approaches to choose and combine the most appropriate solutions and implementations (cloud services, infrastructure resources, deployment scripts, etc.) to establish a particular environment, which depends on individual requirements: a development environment typically aims to be lightweight with minimum overhead, e.g., hosting the entire application stack on a single virtual machine, whereas a production environment has to be highly available and elastic, e.g., comprising a cluster of virtual servers.

Considering performance engineering in the DevOps world, diverse performance aspects of applications are investigated, for example, in the form of an automated performance testing stage as part of a continuous delivery pipeline. However, beside the actually delivered application, its associated continuous delivery pipeline comprising diverse stages and application environments represents a software system on its own. That system is not static but must evolve to deliver new iterations of an evolving application. If, for instance, new kinds of components or technologies are added to the application stack, at least some of the involved application environments need to be updated.
Consequently, performance engineering in the DevOps world is not limited to evolving applications, but also needs to be considered when maintaining continuous delivery pipelines as separately evolving software systems. Beside reliability, a key performance aspect of such pipelines is their speed, i.e., the time it takes from committing a change to the code repository until it is fully tested, packaged, and ready to be put into production. This is the foundation for providing fast and immediate feedback loops to continuously improve the delivered application.

Technically, different approaches can be applied to improve a pipeline's performance. For example, the execution of tests can be split, so that an initial test stage only runs tests that are fast, while following test stages run tests that take longer; this accelerates feedback loops and saves resources by failing as fast as possible. Moreover, binary artifacts such as container images should only be built once and then stored in artifact repositories such as a Docker container registry to be reused in following stages of the pipeline. Another optimization approach could be to consolidate application environments, e.g., to run the build and unit test stages in the same environment. While this helps to lower resource consumption, the degree of isolation is reduced. As a result, this approach cannot be applied if isolation is key (e.g., performance testing vs. production). Another performance aspect is the provisioning time of underlying application environments, e.g., the time it takes to make the unit test environment available to run the unit tests based on it. This could potentially be optimized by reusing environments instead of completely provisioning them for each pipeline run. Performance metrics must be defined for each pipeline, e.g., to measure and record speed and resource consumption of pipelines. This also helps to identify performance regressions of pipelines.
Further approaches to optimize the performance of continuous delivery pipelines may be identified and discussed as part of these research efforts. An initial goal is to come up with a set of patterns and best practices that can be applied to improve the performance of continuous delivery pipelines.



\abstracttitle{Towards Application-aware Cloud Provisioning for Enterprise Applications}
\abstractauthor[Felix Willnecker]{Felix Willnecker (fortiss, Germany, willnecker@fortiss.org)}

\license


Enterprise applications are typically implemented as distributed systems composed of several independent components or services \cite{DBLP:conf/qosa/WillneckerK16}. Modern development paradigms allow to develop such applications using elastic infrastructure as deployment targets. These enterprise applications are designed to scale-out if load peaks occur \cite{DBLP:conf/wosp/Brebner12}. Load peaks are balanced by spawning new virtual servers or containers in an elastic infrastructure [3]. Current strategies are reactive and thus spawn new instances late and often keep this instances for hours after the load peak occurred \cite{DBLP:conf/icse/HerbstKWG15}. In such cases, the infrastructure is over-provisioned for most of the time. Over-provisioned deployments can cause unnecessary costs, as infrastructure providers such as Amazon Web Services (AWS) offer their services based on uptime, load and/or traffic. Better deployment strategies can decrease the runtime costs of enterprise applications by re-using already provisioned server instances or selecting server instances that fit the current demand.

Elastic cloud infrastructures automatically scale deployments based on the load and eliminate the need for sophisticated deployment topologies \cite{DBLP:conf/wosp/Brebner12}. The infrastructures leverage the effect of virtualization and over-commit their hardware. Thus, they create more virtual servers as available hardware would allow. However, the benefits of this over-commitment are solely taken by the infrastructure providers, especially when the virtual servers are poorly utilized. Moreover, if one virtualized server's resources exceed, another instance spawns which increase the costs. Resources of other virtual servers of the same enterprise application are usually not considered to contain load peaks and reduce total runtime costs.

We propose an application-aware cloud provisioning based on a model based deployment topology optimizer \cite{DBLP:conf/qosa/WillneckerK16}. We introduced a deployment topology optimizer, which selects an optimized topology for a fixed workload. The optimizer takes available and unused resources as well as costs and performance into account. In combination with workload prediction and performance models, this optimizer can reorganize distributed enterprise applications based on expected load and already provisioned infrastructure. Thus, reducing the number of instances, which reduces costs and shares the savings of virtual servers in elastic infrastructures with the operator of the enterprise applications.



\section{Working Groups}

\abstracttitle{Performance Engineering Challenges for Microservices}
\abstractauthor[%
Robert Heinrich, 
Andre van Hoorn, 
Holger Knoche, 
Fei Li, 
Ellen Lwakatare, 
Claus Pahl, 
Stefan Schulte, 
Johannes Wettinger
]{%
Robert Heinrich (Karlsruhe Institute of Technology, Germany)\\
And\'{e} van Hoorn (University of Stuttgart, Germany)\\
Holger Knoche (Christian-Albrechts-Universit\"at zu Kiel, Germany)\\
Fei Li (Siemens Corporate Technology, Austria)\\
Lucy Ellen Lwakatare (University of Oulu, Finland)\\
Claus Pahl (Free University of Bozen-Bolzano, Italy)\\
Stefan Schulte (TU Vienna, Austria)\\
Johannes Wettinger (University of Stuttgart, Germany)\\
}
\license


\subsubsection{Discussed Problems}

Microservices are an emerging architectural style, complementing approaches like DevOps and continuous delivery in terms of software architecture. 
To some degree microservices resemble concepts and technologies from other established styles---particularly, service-oriented architectures (SOA). 
The goal of this breakout group was to discuss challenges and possible approaches in performance engineering for microservices. 

We started by identifying differences between microservices and SOAs to see what architectural characteristics stop us from just reusing existing performance engineering concepts. 
 Identified differences include 
i) the technology and protocol stack (e.g., heavy-weight middleware vs.\ light-weight REST-based communication), 
ii) the purpose (e.g., integration of systems vs.\ system decomposition), and
iii) deployment models and scalability. 

Testing, monitoring, and modeling were identified and discussed as primary research areas in performance engineering for microservices. In practice, additional challenges are imposed when existing legacy applications are migrated into a microservice architecture. However, this topic was excluded from the discussion. 

\subsubsection{Possible Approaches}

The core results from the discussion about performance testing, monitoring, and modeling for microservices were:  

\begin{itemize}
	\item \emph{Testing.} 
Early-stage testing 
for microservices follows common software engineering practices. 
More comprehensive testing (e.g., integration/system tests) is compensated or replaced by fine-grained monitoring of production environments. 
%
Performance testing gets simplified in the first place, i.e., the performance of each microservice (typically deployed as a single container) can be measured and monitored in isolation. The continuous delivery practice (e.g., multiple deployments per day) does not allow time-consuming performance testing of the entire application before each single deployment. 
Once a microservice is put into operation and used, the monitored performance data (e.g., using established application performance monitoring (APM) approaches) can in turn be utilized to devise and improve performance regression testing. 

	\item \emph{Monitoring.}
State-of-the-art APM tools 
support the collection of various measures of the entire system stack. 
In microservice architectures, basically the same techniques for data collection can be used as in traditional architectures. 
A technical instrumentation challenge is imposed by the microservice characteristic of polyglot technology stacks, particularly involving the use of  emerging programming paradigms and languages (e.g., Scala).  
In microservice architectures, additional measures are of interest in order to monitor specific architectural patterns at runtime. Examples include the state of resilience mechanisms, such as circuit breakers. 
%
In microservice architectures, it becomes difficult to determine a normal behavior used for anomaly detection. The reason is that due to the frequent changes (updates of microservices, scaling actions, virtualization) no ``steady-state'' exists. Existing techniques may raise many false alarms. 


	\item \emph{Modeling.}
Component-based performance modeling and prediction can be seen as the starting point from which techniques for microservices can be reused. Application-level modeling used to be the focus in traditional component-based performance modeling. Good results could have been obtained by abstracting from the middleware. An example here is work in the Palladio context.

Another perspective is the platform/infrastructure level. Microservice platforms turn out to be more complex and the behaviour of the platform is an influencing factor in the overall setting. Thus, the need emerges to explicitly represent the platform with its different layers (e.g., physical machines, VM, Docker, cluster) on which the application is placed.
Creating models by hand is not an option anymore due to, firstly, the inherent complexity of the platform and, secondly, due to the frequent changes in the environment.
To learn the model in an environment that is bound to change becomes a challenge. Machine learning and model extraction techniques need to be studied to find a solution for the determination and updating of complex, changing models. Work in the context of models@runtime can provide input here.
These models can feed into a closed adaptation loop where for instance auto-scaling techniques are applied. Performance models need to guide monitoring data analysis and decision making (MAPE loop). Controllers must cope with the elasticity rules. New processing policies must be specified.


\end{itemize}

\subsubsection{Conclusions}

Despite the obvious importance of a sufficient level of performance, there is still a lack of performance engineering approaches explicitly taking into account the particularities of microservices. So far, performance engineering for microservices has not gained attraction in the relevant communities and the identified challenges impose a number of interesting research questions and directions. Meanwhile, the results of the breakout group have been summarized in a workshop contribution \cite{HeinrichvanHoornKnocheLiLwakatarePahlSchulteWettinger2017PerformanceEngineeringForMicroservicesResearchChallengesAndDirections}. 

\abstracttitle{Performance Test Prioritization}
\abstractauthor[%
Georgiana Copil, Philipp Leitner, Ingo Weber, Felix Willnecker
]{%
Georgiana Copil (TU Vienna, Austria)\\
Philipp Leitner (University of Zurich, Switzerland)\\ %
Ingo Weber (Data61, CSIRO, Australia)\\
Felix Willnecker (fortiss, Germany)
}
\license

\subsubsection{Discussed Problems}
Performance testing is currently not integrated into the Continuous Delivery pipeline. This is mainly on account of the multitude of performance tests that need to be executed on a high number of system states, resulting in a vast amount of testing configurations. Several approaches, such as BlazeMeter\footnote{\url{https://www.blazemeter.com}}, LoadImpact\footnote{\url{ https://loadimpact.com}}, or RadView\footnote{\url{http://www.radview.com}} offer PaaS solutions for performance testing under production-like configurations, with possibilities to integrate them into normal continuous delivery pipelines (e.g., BlazeMeter Jenkins plugin). 

\begin{figure}[h]
	\centering
	\includegraphics[width=0.5\columnwidth]{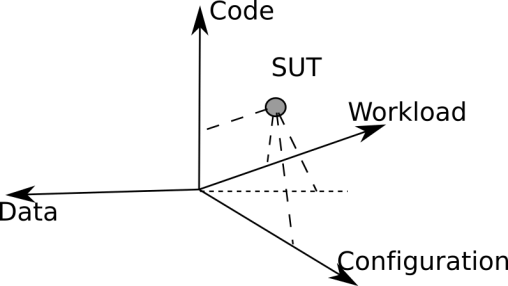}
	\caption{Factors of influence for performance testing.}
	\label{fig:factors}
\end{figure}

However, the cost and time needed for executing performance tests can be prohibitive, leading to these tests being absent from most continuous delivery pipelines. 
Figure~\ref{fig:factors} shows the defining dimensions influencing runtime performance, which characterize the System Under Test: (i) the code that is being tested, (ii) the workload, (iii) the test data, and (iv) the configuration. The latter includes all software system and hardware related configurations, defining the environment in which the system under test will be running. 

\subsubsection{Possible Approach}

In the breakout group, we have identified test case prioritization as a promising approach to deal with these issues: prioritizing which of the performance tests need to be run in order to determine most important regressions, and excluding tests that determine the same performance regression. Executing the performance tests covering the most important performance regressions can lead to improved testing time and reduced cost, or can support software developers with limited time or limited budget in discovering possible performance regressions. 

The core idea of this work is to determine which of a number of microbenchmarks are indicative of true runtime  performance of an application. After obtaining that knowledge, we can use it for prioritization of performance microbenchmarks to make the most of a given time budget for performance testing, as well as for utilizing the performance measurements obtained as a side effect of regular unit testing for performance regression testing.

\begin{figure}[h]
	\centering
	\includegraphics[width=0.75\columnwidth]{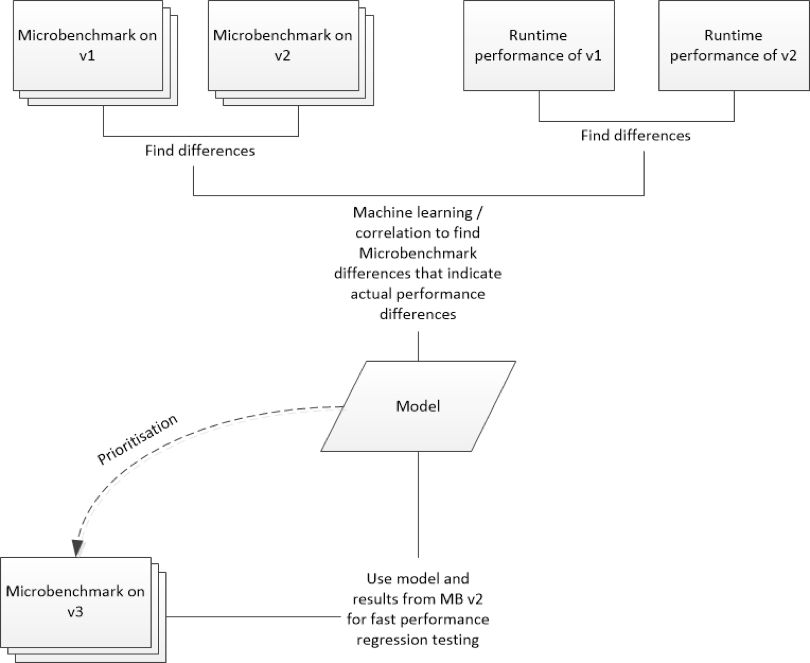}
	\caption{Overview of the proposed machine learning based performance test prioritization approach.}
	\label{fig:approach}
\end{figure}

The means by which we plan to obtain this information is machine learning or statistical techniques such as correlation, as shown in Figure~\ref{fig:approach}. Specifically, we plan to use these approaches to determine which differences of performance in microbenchmarks from one version of the software to the next are indicative of actual performance differences between these versions.

\subsubsection{Conclusions}

We plan to use machine learning and statistical correlation analysis to identify particularly ``promising'' performance tests to quickly execute as part of a Continuous Delivery pipeline for fast performance feedback. We hope that this work will lead not only to useful methods for microbenchmark evaluation, but also to immediately useful tools that can be integrated into CI/CD pipelines (e.g., Jenkins plugins). Finally, we expect to improve our understanding of how to write ``expressive'' microbenchmarks in the first place.

\abstracttitle{Uncertainty in a Performance-Aware DevOps Context}
\abstractauthor[%
Markus Borg, J\"urgen Cito, Pooyan Jamshidi, Zhen Ming (Jack) Jiang, and Catia Trubiani %
]{%
Markus Borg (RISE SICS AB, Lund, Sweden)\\
J\"urgen Cito (University of Zurich, Switzerland)\\
Pooyan Jamshidi (Imperial College London, United Kingdom)\\%
Zhen Ming (Jack) Jiang (York University, Toronto, Canada)\\
Catia Trubiani (Gran Sasso Science Institute, Italy)
}
\license


Uncertainty is a very relevant challenge to performance-aware DevOps.  
Performance measurements and predictions are essential in DevOps solutions to the extent that many different tools rely on the performance measurements and predictions \cite{XX-SPEC-RG-2015-001-DevOpsPerformanceResearchAgenda}.
Various techniques to evaluate the performance properties of software systems in early stages of development exist, e.g., based on architectural models enriched by performance-relevant information. However, in early stages, various parameters of the systems are uncertain, e.g., regarding implementation details and environment in which the software systems are deployed. This imposes a challenge on early performance prediction. In the software performance domain, uncertainties concern, for example, the usage profile (including workload intensity, navigational profiles, and input data), resource demand characteristics of software services, and properties of the deployment and execution environment (including hardware) on which the software will be deployed. 

In the break out group, we discussed the uncertainty challenges in the context of DevOps and we based our general definition of uncertainty as \cite{walker2003defining}: ``any deviation from the unachievable ideal of completely deterministic knowledge of the relevant system''. Such deviations can lead to an overall ``lack of confidence'' in the obtained predictions based on the monitoring data that they might be ``incomplete, blurred, inaccurate, unreliable, inconclusive, or potentially false''.
To make informed decisions, DevOps teams need to be aware of uncertainties in the whole DevOps life-cycle to be able to interpret data, models, and results accordingly. We have identified sources of uncertainty in a typical performance-aware DevOps scenario.

\subsubsection{Discussed Problems}
Goal of this break out group was to identify the sources of uncertainty due to the application of DevOps in the performance evaluation of software systems. In fact, there is an obvious trade-off in the performance evaluation of early model abstractions where the amount of information is limited, and late performance monitoring on running artifacts where the results are more accurate but some constraints have been added. We were discussing our experiences on case studies showing different sources of uncertainty that span on multiple characteristics, such as the structural, behavioural, and deployment aspects of a software system and it emerged that performance results are heavily affected by these aspects. 

\subsubsection{Possible Approaches}
To make informed decisions, DevOps teams need to be aware of uncertainties in the whole DevOps life-cycle to be able to interpret data, models, and results accordingly. To provide support in this direction, possible approaches have to: (i) identify sources of uncertainty in a typical performance-aware DevOps scenario; (ii) elaborate how these uncertainties manifest in input data, design models and operational results; (iii) make suggestions as how to interpret knowledge (i.e., input data, design models and operational results) given these sources of uncertainty. In this way it is possible to figure out the performance trend of the system exposed to such uncertainties. 

\subsubsection{Conclusions}
Our conclusions were that it is relevant to bring the sources of uncertainty up-front in the performance-aware DevOps process to support developers in the interpretation of performance evaluation results. Indeed, more research is needed in this direction; we plan to investigate this topic further to investigate how the explicit specification of uncertainties benefit the performance analysis process.

\abstracttitle{How Can We Facilitate Feedback from Operations to Development?}
\abstractauthor[%
Markus Borg, J\"urgen Cito, Fei Li, Lucy Ellen Lwakatare, Johannes Wettinger%
]{%
Markus Borg (RISE SICS AB, Lund, Sweden)\\
J\"urgen Cito (University of Zurich, Switzerland)\\
Fei Li (Siemens Corporate Technology, Austria)\\
Lucy Ellen Lwakatare (University of Oulu, Finland)\\
Johannes Wettinger (University of Stuttgart, Germany)
}
\license 


\subsubsection{Discussed Problems}
Feedback from operations is important to drive informed decisions especially in modern software development approaches of fast release cycles. When software is operated in production, it produces a plethora of data that ranges from log messages emitted by the developer from within the code to performance metrics observed by monitoring tools. All this data gathered at runtime serves as valuable \emph{feedback} to various stakeholders to improve the software itself and the process overall.  However, there are some challenges organizations face when attempting to facilitate proper feedback channels between operations and the rest of the software development life-cycle. Some of these challenges include: \textit{organizational size, nature of business, presentation of feedback and other technical challenges}. We discussed these challenges in more detail and looked at possible solutions.

\subsubsection{Possible Approaches}
\begin{figure}[h]
	\centering
	\includegraphics[width=0.95\columnwidth]{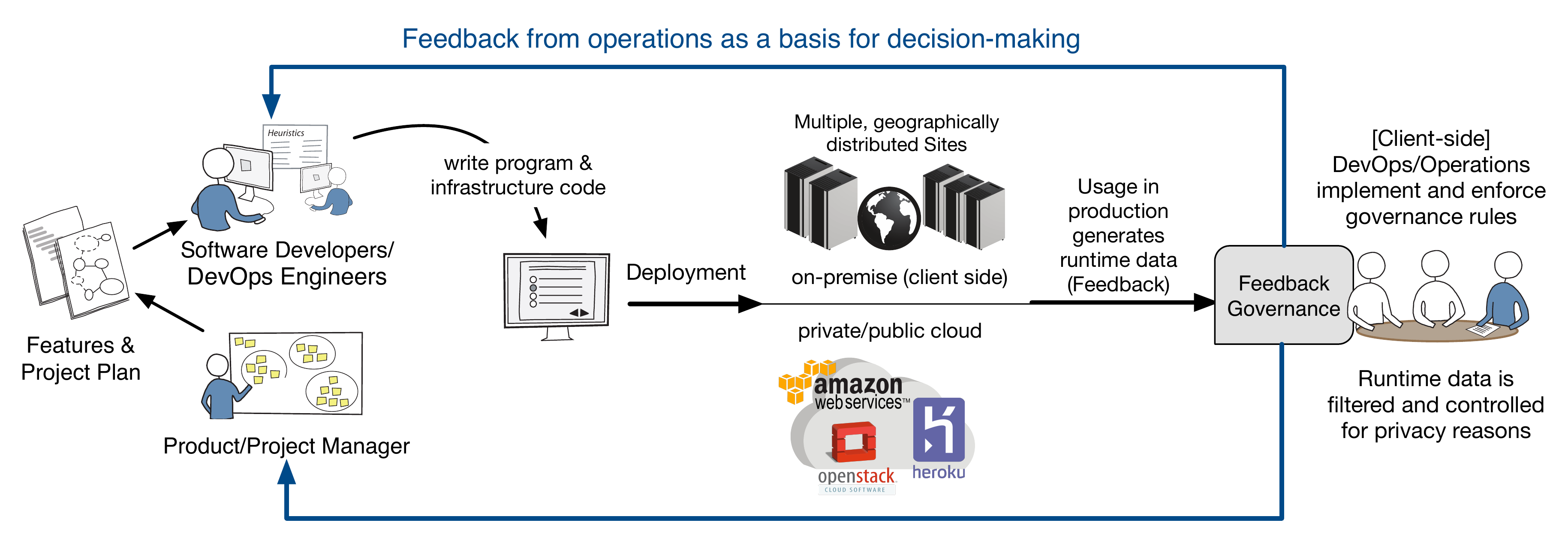}
	\caption{Feedback process involving multiple stakeholders and organization boundaries.}
	\label{fig:process}
\end{figure}

Figure \ref{fig:process} illustrates a generic \emph{feedback process} that attempts to abstract the process of retrieving feedback from operations across different organizational boundaries. We derived the process from informally discussing industrial use cases to ensure that the process covers concerns and needs of different company structures. 

Taking this generalized feedback process as a basis for discussion, we formulate the following considerations:
\begin{itemize}
        \item \emph{Role of Deployment for Feedback}.  The feedback process is initially kicked-off with deployment of software to make it available to end-users. Deployment can range from an automated, continuous delivery/deployment process to releasing a software unit that requires more complex (often manual) processes to roll out.
In the former case, it stays within a company’s own organizational boundaries (public/private cloud or data center). Product development together with DevOps/operations engineers from the same organization are responsible for the operability. In the latter case, it is delivered through consultants/solution architects as on-premise software. 
The complexity to enable a proper feedback process thus depends on the complexity of the deployment pipeline. 
        \item \emph{Feedback Governance}. There needs to be control over which kind of operations feedback is available to which kind of stakeholder. This kind of governance should explicitly provide high-level rules on how data is handled either in organizations or as a cross-organizational concern. These rules are then implemented and enforced by the DevOps/operations engineers by filtering and controlling runtime data. The consequence of this part of the process is that privacy is being enforced and product development only has access to data that exhibits no threat of violations or non-compliance. 
        \item \emph{Closing the Feedback Loop: Decision-making in Development}. Eventually, once operations data passes through governance it becomes valuable feedback to stakeholders in product development. Figure \ref{fig:process} illustrates two broad examples of stakeholders benefiting from runtime feedback. Product/project managers can now use runtime feedback to better plan their features and optimize their project plan. Software developers and DevOps engineers have a full picture of how users experience their software (e.g., performance metrics, usage counters) and can tweak program and infrastructure code to improve the overall experience. Here, the feedback loop starts again with deploying changes to software that were informed by better decisions through runtime feedback.
 \end{itemize}       
\subsubsection{Conclusions}

Runtime aspects of software observed and collected as metrics and events can provide valueable feedback in the development cycle of software. We discussed different considerations of establishing a feedback process that depends on the complexity of the existing deployment structures and requires its own feedback governance to be established (especially in larger organizations). We plan to further investigate industrial case studies to identify the challenges when establishing such a feedback process.

\abstracttitle{Performance Engineering for Blockchain-based Applications}
\abstractauthor[%
Philipp Leitner,
Stefan Schulte,
Ingo Weber
]{%
Philipp Leitner (University of Zurich, Switzerland)\\
Stefan Schulte (TU Vienna, Austria)\\
Ingo Weber (Data61, CSIRO, Australia)\\
}
\license
\subsubsection{Discussed Problems}
Blockchain is often named as a way to realize trust between anonymous parties without the need of a trusted third party. Instead, trust is established as an emergent property of the blockchain technology, i.e., a distributed ledger which stores transactions in a permanent and indisputable way~\cite{tschorsch16}. Blockchains are not maintained by a single organization, but hosted and enacted in a peer-to-peer manner. Blockchains can be used in arbitrary applications in order to provide provenance of data, e.g., about business transactions. For instance, blockchains have been applied in order to execute business processes as smart contracts~\cite{weber16}, thus documenting the process execution in a blockchain and using smart contract features to ensure that the participants in the collaborative process do not deviate from the agreed-upon process model.

Within this breakout group, our goal was to identify performance engineering issues for blockchain-based applications and to think about respective solution approaches. Indeed, there has only been limited discussion on how the usage of blockchain technologies influences application design \cite{xu16}. A particular aspect that has not been discussed in detail are timing issues. New blocks in blockchains are usually issued with a particular time distance between two blocks, i.e., the so-called \emph{interblock time}. For instance, in the Bitcoin blockchain~\cite{nakamoto08}, the median interblock time is set to 10 minutes; in the Ethereum blockchain~\cite{ethereum14}, it is set to about 13 seconds. Furthermore, there is a lot of variance: individual interblock times for Bitcoin can easily exceed an entire hour. Applications which want to apply the blockchain need to cope with this and accept it as given. 

In the following subsection, we discuss how these timing issues as well multi-step transactions lead to (performance) uncertainty in blockchain-based applications. Afterwards, we propose how to overcome these issues.


\subsubsection{Possible Approaches}
\begin{itemize}
	\item \emph{Performance Uncertainty.} A core performance challenge in blockchain-based applications is caused by the peer-to-peer nature of the network, as well as the cryptographic properties of the different blockchain protocols. Concretely, we have discussed two sources of performance variability and therefore uncertainty in our breakout group:
	\begin{itemize}
		\item \emph{Uncertainty with regards to the interblock time.} While protocols such as Ethereum aim to normalize this time to a well-defined value in the median, substantial outliers exist. For instance, for Bitcoin, in a 10-minute interval the next block is found with a probability of 63\%, but roughly 5\% of the interblock times are above 30 minutes\footnote{Source: \url{https://en.bitcoin.it/wiki/Confirmation}}. Ethereum suffers from similar variance. As can be seen, interblock times are uncertain, and may vary to a very large extent~\cite{yasaweerasinghelage17}.
        
		\item \emph{Uncertainty with regards to the confirmation of a block.} A transaction being part of a single block does not guarantee that the transaction will remain part of the chain. The transaction may still ``fall off'' if the blockchain decides that the block the transaction is part of becomes deprecated, because of a fork of the chain. However, the more follow-up blocks (also called confirmation blocks) have been added to the blockchain after the transaction has been added, the more unlikely this becomes. In consequence, the probability of a transaction actually being and staying part of a blockchain increases with time.
	\end{itemize}
	In order to overcome the issues arising because of performance uncertainty, it is first necessary to empirically examine the blockchain's behavior. This requires observing the blockchain networks and to statistically model interblock times as well as the likelihood and frequency of forks. Once this has been done, the likelihood of forks can be modeled using the means of Markov chains, or, assuming that the probability remains constant, as a simple conditional probability. A first short paper in this direction has been published after the seminar~\cite{yasaweerasinghelage17}.
	
	\item \emph{Uncertainty in Multi-Step Transactions.} Uncertainty may also arise from applications which include multi-step transactions.
    If we assume transactions to not only be simple one-shot interactions with the blockchain, but rather multi-step (business) processes~\cite{MWA+17}, uncertainties become even more problematic. For instance, in a business process, process state transitions are triggered by events. Each event can be stored in a blockchain. However, due to the uncertainties mentioned above, a process participant may be required to wait until an event has $n$ confirmation blocks on the blockchain before she considers an event to ``have happened for sure''. This leads to considerable delays in state transition times if every transition needs to be delayed until enough time has passed and confirmation blocks have been accumulated to reach sufficient certainty.
    
    Alternatively, a process participant may decide that she ``accepts'' a state transition after a shorter period of time in order to speed up the process, for instance because she can observe the effects of the change in the physical world, or because the specific state transition is not crucial to her.
	
	However, this leads to interesting challenges, both from a business process modeling and execution perspective:
    \begin{itemize}
		\item From a \emph{modeling} perspective, it raises the question how a process participant would model what level of certainty she requires for each transition. 
        \item From an \emph{execution} perspective, the blockchain-based engine needs to be aware of the possibility that it is ``wrong'' about the current state of the process for any given instance. Note that this may have domino effects, as participants may have already taken follow-up actions based on previous state information that may not be valid anymore after an update. A blockchain-based process modeling language and engine should have means to specify and execute suitable compensations for such cases, which bring the process back into a consistent state.
	\end{itemize}
\end{itemize}

\subsubsection{Conclusions}
To the best of our knowledge, despite the rapidly increasing application of blockchain technology by the research community and industry, performance engineering considerations around the use of blockchains in application design and development have not gained much attention yet.

As concrete follow-up steps of the discussions within the breakout group, 
we will carry out the needed empirical studies and take the results from these studies into account in order to devise uncertainty-aware, blockchain-based applications and suitable performance engineering approaches. 

\abstracttitle{Implications of DevOps and Self-Adaptivity}
\abstractauthor[%
Georgiana Copil, Pooyan Jamshidi, Cristian Klein, Claus Pahl%
]{%
Georgiana Copil (TU Vienna, Austria)\\
Pooyan Jamshidi (Imperial College London, United Kingdom)\\
Cristian Klein (Umea University, Sweden)\\
Claus Pahl (Free University of Bozen-Bolzano, Italy)%
}
\license


\subsubsection{Discussed Problems}

Self-adaptivity is a software engineering concept aiming to reduce runtime uncertainty at design time by designing the application to adapt to runtime changes. For example, the uncertainty in the number of users accessing a particular application can be reduced by designing the application with auto-scaling capabilities. At runtime, the application will adapt to the actual number of users, by acquiring and releasing computing capacity, usually presented as virtual machine instances, or enabling/disabling optional features of the system.

We discussed two possible relationships between self-adaptivity and DevOps: (a) adding DevOps to a self-adaptive application (DevOps4SA) and (b) adding self-adaptivity to a DevOps pipeline (SADevOps). For both topics, we discussed motivation, challenges and opportunities.

\subsubsection{DevOps4SA}

Using DevOps for delivering a self-adaptive application is motivated by bringing the benefits of evolving both the application and its controller in a reliable manner. However, this would also bring more challenges. First, there would be more sources of uncertainty, bugs and regressions, which could come from the application, the controller or the interaction between the two. Second, controllers and applications are often tightly coupled. Hence evolving both of them would be challenging. Finally, due to the frequent nature of deployments in the DevOps culture, the controller would have shorter learning periods between consecutive changes to the application. Indeed, new code added to the application could invalidate the model learned by the controller. The model learning can be done in an incremental fashion.

We also discussed opportunities brought by {\sf DevOps4SA}. First, the DevOps pipeline could be accelerated by reducing testing time, since uncertainties are compensated for at runtime. Second, the controller could reuse the learned model from previously deployed version of the application, hence not needing to relearn a model of the application from scratch.

\subsubsection{SA4DevOps}

One of the main ideas of DevOps is to use measurements from operations as feedback for development. Making the DevOps pipeline self-adaptive would allow taking automated low-level decisions, freeing up humans to deal with higher-level decision. The system would be composed of two loops. The human-in-the-loop would focus on delivering value and deal with long-term predictions, aiming to evolve requirements, design and implementation of the application, as promoted by the BizDevOps concept. The controller-in-the-loop would focus on dull and error-prone processes and deal with short-term predictions, aiming to evolve the configuration. The Dev and Ops would essentially acts as supervisors for the self-adaptive controller.

The discussion then naturally steered towards what kind of DevOps activities could be taken over by self-adaptive loops:

\begin{itemize}
\item\textbf{Remap for Performance:} A controller could identify hotspots during operation --- for example, two components that frequently feature high CPU utilization simultaneously --- and add anti-co-location constraints to be taken into account during the next deployment.
\item\textbf{Remap for Resilience:} Similarly to the example above, a controller could identify hotspots that appeared after fault injection and add anti-co-location constraints.
\item\textbf{Fault Injection:} Based on the log of source code changes, a controller could inject faults targeted at application components that have recently changed, since these are more likely to be plagued by bugs.
\item\textbf{Monitoring:} Similarly to the example above, a controller could increase the monitoring frequency of recently changed application components in order to get more information at a certain time period.
\item\textbf{Anomaly Detection:} Taking as input source code changes and test results, a controller could change alarm thresholds and whether to alert developer or operators. A recently changed component is more likely to feature a bug within developer's responsibility, whereas a stable component is more likely to feature erroneous behaviour due to changes in operations.
\item\textbf{Tracing:} Tracing large-scale distributed systems can be very expensive, therefore, tracing is usually turned off or set to a low level of detail. Taking as input source code changes and test results, a controller could enable tracing or increase the level of detail for application components that recently changed.
\end{itemize}

Most of the above topics consider code change in the system as enviromental uncertainty that requires an appropriate reaction from the down-the-line DevOps toolchain. 

\subsubsection{Conclusions}

We found many promising research directions arising from the interaction between DevOps and self-adaptivity. DevOps can enable co-evolution of controllers and the controlled system. On the other hand, self-adaptivity can provide opportunities to automate manual and labor intensive tasks in the DevOps pipeline.

\abstracttitle{A Systematic Process for Performance Antipattern Detection and Resolution in DevOps based on Operational Data and Load Testing}
\abstractauthor[%
Alberto Avritzer, Andr\'{e} van Hoorn, Catia Trubiani, Holger Knoche%
]{%
Alberto Avritzer (Sonatype, USA)\\
Andr\'{e} van Hoorn (University of Stuttgart, Germany)\\
Catia Trubiani (Gran Sasso Science Institute, Italy)\\
Holger Knoche (Christian-Albrechts-Universit\"at zu Kiel, Germany)%
}
\license


\subsubsection{Discussed Problems}
We discussed the problem of providing the performance assessment of complex software systems since such systems are subject to many variabilities, such as workload fluctuation and resource availability. The main issue is that such variabilities may introduce flaws that affect the system quality and generate negative consequences, such as delays and failures. Hence, it is necessary to put in place some methodologies that allow to recognize performance flaws and generate software refactorings able to overcome such flaws.

\subsubsection{Possible Approaches}
We are investigating an approach based on load testing and profiling data, to identify performance flaws and generate software refactorings to developers, thus to support them in the interpretation of performance measurements. These activities are supported by performance antipatterns that are well known to document common development mistakes leading to performance flaws as well as their solutions. To this end, we are investigating a real-world case study provided by an innovative company in the open-source domain.

\subsubsection{Conclusions}
We found that the analysis of load testing and profiling data is not trivial, and the domain expert knowledge is fundamental to progress on such activity. The automation of software refactorings is feasible up to a certain extent, there are some peculiarities that cannot be handled by tools since they are not machine-processable due to their nature.

\abstracttitle{Models@DevOps}
\abstractauthor[%
Robert Heinrich, J\"urgen Walter, Felix Willnecker%
]{%
Robert Heinrich (Karlsruhe Institute of Technology, Germany)\\
J\"urgen Walter (Julius Maximilians Universit\"at W\"urzburg, Germany)\\
Felix Willnecker (fortiss, Germany)%
}
\license


We see a culture clash of DevOps and traditional performance modelling (cf. Table \ref{tab:clash}). Performance modeling in traditional Software Performance Engineering is characterized by manual efforts for creating and parameterizing performance models during design time. Usually there is no continuous performance analysis and design optimization. 
%
In contrast, DevOps practices are characterized by short, fast automated release cycles, agile processes, short feedback-loops and continuous executions of elaborated, automated, and staged test chains. In dynamic systems, performance goals need to be validated and maintained continuously. 

\begin{table}[t]
  \centering
  \caption{Culture Clash Between MDSD and Modeling}
  \label{tab:clash}
  \begin{tabular}{p{6.5cm}|p{6.5cm}}
    \toprule
    MDSD & DevOps \\ \midrule
   \begin{itemize}
\item Create design-time model manually
\item No continuous performance analysis
\item No ongoing design optimization
\end{itemize}
&
\begin{itemize}
\item Short/fast automated release cycles
\item Agile processes, short feedback-loops
\item Continuous execution of elaborated, automated, and staged test chain
\end{itemize}
    \\
   \bottomrule
  \end{tabular}
\end{table}


\subsubsection{Discussed Problems}
Due to differences the question comes up: ``Do we still need performance models in an agile DevOps world?''
Models are not yet common industrial practice. However, there are open DevOps challenges that can be addressed using performance models. Further, we see DevOps as an enabler for performance models.
The need is also driven by new application scenarios, e.g., Internet of Tings (IOT) causes a huge number of different devices for which one cannot built a complete test bed.

\paragraph*{DevOps Challenges and Model-based Solution}
The results of our discussion are presented in Table \ref{tab:model_answers}

\begin{table}[t]
  \centering
  \caption{DevOps Challenges and Model Answers}
  \label{tab:model_answers}
  \begin{tabular}{p{6.5cm}|p{6.5cm}}
    \toprule
    Test execution time delays feedback. System analysis only in real time possible & 
    Models analysis instead of measurement-based testing \\
     \midrule
    Amount of monitoring data too complex for human mind & automatic processing and aggregation necessary. Model-based representation of system \\
     \midrule
    Monitoring overhead (thus costs) high & Use simulations instead of system monitoring \\
     \midrule
   Common language between developers and operators are missing & Specific views for operator and developer on the same model\\
   \bottomrule
  \end{tabular}
\end{table}

\subsubsection{Possible Approaches}
\begin{figure}[h]
	\centering
	\includegraphics[width=0.75\columnwidth]{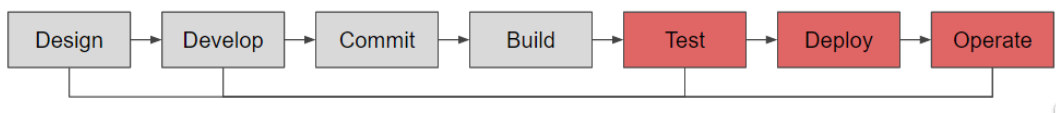}
	\caption{DevOps Pipeline.}
	\label{fig:pipeline}
\end{figure}

\paragraph*{Pipeline Integration of Performance Models }
We discussed where models can be used in the DevOps Pipeline (cf. Figure \ref{fig:pipeline}).

\medskip

\noindent \textbf{Fully automated}
\begin{itemize}
\item Regression analysis
\item Runtime adaptation / self adaptation
\item Recovery models
\item Resilience analysis
\item Forecast-based decisions
\end{itemize}

\noindent \textbf{Human in the loop}
\begin{itemize}
\item Design space exploration / Decision making
\item Performance models as simplified view on APM data
\item Parallel evaluation of multiple design alternatives 
\end{itemize}

Performance models and prediction can be integrated to automate certain decisions but also to assist developers, designers, architects or operators. Performance models can be used to explore a design space by evaluating a number of design decisions by alternating an extracted model and simulate the effects. Such simulations can be executed in parallel and thus evaluate multiple system variations at the same time. Performance models are also suitable for communication as they represent a simplified view on the system but are at the same time based on detailed measurements. 
Such models can be used for fully automated regression analysis. Integrated in a continuous delivery pipeline, such evaluations can be conducted where classical performance tests take too long. The resilience and recovery capabilities of a system model can be conducted and evaluate recent changes in regards of fault tolerance of the system. Forecasts and analysis on these models and predictions allow to improve self-adaption of systems and especially with forecasts to speed-up scaling decisions. A resource shortage can be predicted before it occurs and replica systems can thus be spawned and warmed up before the actual shortage occurs. 

\subsubsection{Conclusions}
\paragraph*{DevOps as Door Opener for Performance Models} 
Hypothesis: DevOps is an enabler for Performance Modelling. 
\begin{itemize}
	\item Especially continuous monitoring is an enabler for automatic performance model generation
	\item Model extraction can be placed in the pipeline
	\item Huge amounts of APM data available to use machine learning to significantly improve models
 	\item New use cases and scenarios for performance models
\end{itemize}
Application performance monitoring/management software is nowadays integrated in a vast amount of systems, especially in the DevOps context. This monitoring allows to ensure and enforce software quality with regards to their Service Level Agreements (SLAs) even though fast release cycles change the software quite often. The availability of such monitoring data is an enabler for performance modelling as it allows to generate performance models from these application specific traces. On the other hand, these models can be again applied to the DevOps pipeline to replace and/or enhance performance evaluations that usually take too long to execute in an automated delivery pipeline.

\paragraph*{What Has to Change}
Models did not yet arrive in the DevOps world. Heavy weight processes. In order to address the challenge, we postulate things that have to change \ldots

\begin{itemize}
\item More automation 
\item Automatically test validity of models
\item DevOps provides chance to learn application/changes
\item Accuracy tuning for extraction mechanisms
\item Different granularity levels (views + solving) 
\item Focus on dynamic behavior instead of steady state
\item Models that are feasible for automated processing and comprehensible by humans
\item Partly model updates 
\end{itemize}

\abstracttitle{Performance Testing with a Limited Budget}
\abstractauthor[%
Cor-Paul Bezemer, Lubomír Bulej, Vojtěch Horký, Zhen Ming `Jack' Jiang, Dusica Marijan, Weiyi `Ian' Shang%
]{%
Cor-Paul Bezemer (Queen's University, Canada) \\
Lubomír Bulej (Charles University, Czech Republic)\\
Vojtěch Horký (Charles University, Czech Republic) \\ 
Zhen Ming `Jack' Jiang (York University, Canada) \\
Dusica Marijan (Simula, Norway) \\
Weiyi `Ian' Shang (Concordia University, Canada)%
}
\license

In comparison to functional unit testing, performance testing is still not a widely adopted software development practice. This is likely because there are considerably many barriers that hinder adoption of performance testing.
Measuring and comparing performance on modern platforms is objectively difficult, and great attention must be paid to proper collection and analysis of the results, especially when measuring time quantities with sub-millisecond resolution. A performance unit test is not black and white (which makes it more difficult to interpret), it usually takes much more time (which makes it unsuitable for quick testing), and is more difficult to construct (developers need to know which features need to be tested for performance and submit those to a realistic workload).
In contrast, functional unit tests are considerably easier to create and provide clear benefits that are now understood by many. In addition to providing evidence that the software works as expected, they are used to drive the design and enable design changes that are often necessary to suit the everchanging requirements.
While the understanding of measurement and analysis techniques has improved over the last few years, some of the issues remain, and they are related to the difficulty the developers have with interpreting performance test results, finding the right features to test, and finding representative workloads. However, we expect this to change as more and more teams adopt the DevOps culture, which brings together developers and operators to ensure continuous delivery of high-quality code.
We assume that increased adoption of the DevOps culture will increase the demand for performance testing, and at the same time provide developers with information from ``Ops'' that will guide performance testing activities. This includes information that might have been previously lacking, such as usage data for features, platforms, and configurations, as well as representative workloads, without which developers were reluctant to invest effort into performance testing.
Assuming there is demand for performance tests and that we have access to information from ``Ops'' to guide performance testing, the discussion in this group focused on our ability to conduct performance testing in a timely manner. 

\subsubsection{Discussed Problems}
Performance testing takes a lot of time and often requires dedicated hardware resources. Performance unit tests often measure time intervals in the order of milliseconds or below, and if we want to automatically detect performance differences in the order of 10\% or smaller, we need to use sound statistical methods. This in turn requires steady-state samples from multiple test runs, which also requires spending considerable time (with respect to the duration of the measured operation) on system warm-up. If we focus on large-scale performance or load tests to detect only critical and easy to detect performance regressions (to avoid deploying broken software), we may not need as many samples and test runs, but such tests take long time to setup and execute anyway (e.g., replaying a 24-hour long request log).
In addition to the number and type of performance tests, there are other dimensions that greatly influence the amount of time needed for performance testing. These include different hardware platforms and hardware configurations, multiple software platforms and their configurations, and last, but not least, stream of changes to the software resulting in new versions that need to be tested.
Ideally, we want to fully test each version found in a source code repository. However, just multiplying the number of tests, the required number of test runs, the number of hardware platform configurations, and the number of software platform configurations is bound to produce a number of test executions that we cannot hope to fit into a limited time and cost budget. Therefore we need to schedule tests to make the most of the available budget.

\subsubsection{Possible Approaches}
Selecting the right tests under given time or cost constraints is essentially a planning problem, except that the cost and importance of tests is not completely static. Instead, it is influenced by various inputs. We therefore need a test scheduler that can, for each software or configuration change that may affect performance, maximize the chance to find performance regressions under the given time and cost constraints using the following:

\begin{itemize}
\item The available tests and their duration
\item Testing requests from developers
\item Testing guidance from operators
\item A history of results for a particular test
\end{itemize}

\begin{figure}[h]
	\centering
	\includegraphics[width=0.75\columnwidth]{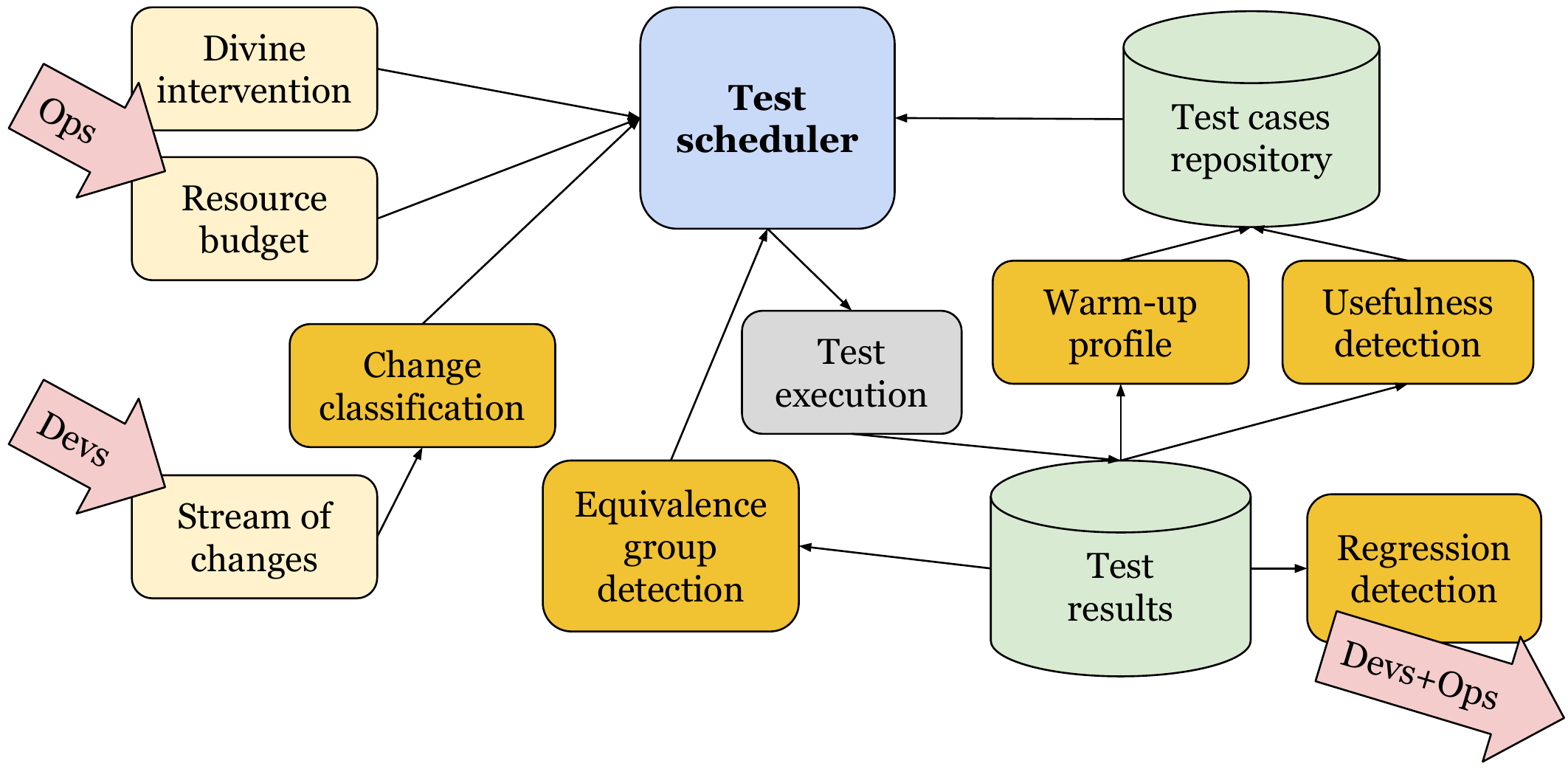}
	\caption{Overview of the proposed architecture for a performance test scheduling system.}
	\label{fig:limited_budget}
\end{figure}

Figure~\ref{fig:limited_budget} shows a coarse architecture of such a test scheduling system. The test scheduler itself is only one of several components, and is intended to make decisions based on information coming from multiple sources. The operators determine the resource budgets as well as the general focus of testing, while developers may influence test selection or test priority in response to actual needs. Besides the input from developers and operators, the activity of the scheduler is triggered by incoming code/configuration changes, and the test selection and prioritization takes into account previously collected information about tests. The latter comes from a repository which is mainly updated by other processes based on test results. The processes related to ``test execution'' and ``regression detection'' are assumed to be well understood, and are not discussed here in detail---the role of these processes is to provide the system with raw data from test execution and to detect performance regressions.
The other processes provide opportunity for optimization,  and include prefiltering of changes, test case prioritization, test case selection, and test case execution time. We now discuss these processes in more depth.

\paragraph{Change Classification}
If we were to run the full battery of changes on each commit observed by the test scheduler, the system would most likely fail to test the system for performance regressions in a timely manner. It is therefore important to understand the nature of observed changes and adapt the test selection to the ``test worthiness'' of a particular change. For example, changes in source code that only contain formatting or comment changes should be ignored, as they are not expected to change the performance of the resulting system.

A more complex change analysis may involve mapping of source code to nodes in a feature model, which can be further annotated with information about tests covering a particular feature (e.g.,~\cite{chen2000case, gotlieb2016}). This may allow the scheduler to only schedule tests related to a particular feature or features that depend on it.
There is a potentially significant body of related work in the software engineering community, related to change impact analysis and subsequent selection of unit tests (e.g., \cite{orso2004empirical,wloka2009junitmx}). However, these works are mostly concerned with functional unit testing, while performance tends to be a cross-cutting concern, so it may be more difficult to isolate the code the performance of which is supposed to be influenced by a code change and map it to a particular performance test.
Nevertheless, the classification of changes for the purpose of test selection or prioritization is an important aspect of the system that determines the general direction of performance testing, and will benefit from operator (and possibly developer) input.
We assume each test to have two kinds of priorities. The first priority is long-term (or static), based on importance of certain features (based on information from Ops) and the history of results of a particular test case (i.e., how often a particular test case identified a performance regression). The second priority is short-term, based on the results of change analysis, on requests from developers for executing a specific test case (to accommodate their gut feelings), and on regression analysis results (so that a change in which a regression was detected can be tested more thoroughly, including surrounding changes in the project’s development history).

\paragraph{Equivalence Group Detection}
Ideally, software features will be covered by multiple targeted performance tests to ensure that changes to the software do not cause performance regressions. However, in many scenarios we envision using performance tests that are not targeted at a specific feature, and instead exercise the system at a more coarse scale. Benchmarks from various benchmark suites can often serve as performance tests, and may be used to track performance and detect performance regressions in a software project.
While coarse-grained tests (benchmarks) are undoubtedly useful, their results for many changes may be correlated, suggesting that only a subset of the tests may need to be executed. Tests that detect the same performance anomalies should therefore be organized into equivalence groups, and the scheduler should only select a subset of tests that covers all equivalence groups. To prevent selection bias, the tests within an equivalence groups should be selected in a round-robin fashion. However, a single test may be potentially a member of multiple equivalence groups, which requires a two-level selection process to ensure that different permutations of tests across all equivalence groups are used to avoid test starvation.

We also assume that the group membership may change over time, and will therefore require periodic re-evaluation, with the period depending on a particular project. This requires the equivalence group evaluation process to be automated as much as possible.

\paragraph{Test Utility Profiling}
While no test that is part of the code base should be entirely avoided by the scheduler (if a test is considered completely useless, it should be removed), the scheduler should attempt to maximize the likelihood of finding performance regressions by scheduling tests with a history of detecting performance regressions more often.
To simplify the scheduler internals, we assume that the information that determines the priority/weight of a particular test case should be available to the scheduler from the test information repository. The long-term weight/priority of a test should reflect its history of detecting performance regressions, and we assume that it can be produced by an independent process that consumes the results of the regression detection process.

If the global information about test history does not provide sufficient information for test prioritization, it may be necessary to associate the history of test results with additional information about the change that triggered the testing, such as code locations or feature model nodes.

\paragraph{Test Warm-up Profiling}
An obvious way to reduce testing time is to reduce the duration of test execution. However, the challenge is to determine for how long a test needs to execute. Tests in which the workloads take milliseconds or less to execute (benchmarks and microbenchmarks) typically repeat the workload many times so that the measured duration of the operation can be processed in a robust fashion using statistical methods. Prior to collecting the measurements, a test harness typically performs a warm-up phase to get rid of various transients associated with the start up of a system or an execution platform, such as the Java Virtual Machine. This needs to be repeated multiple times by executing the workloads in newly spawned processes to take into account changes in performance due to layout of code and data in memory, and thus properly sample the observable variance in the measured data.

Often, the samples from different test runs contribute more information to the estimate of the observable variance than the samples from a single execution. The warm-up phase of a test run then becomes a significant overhead in test execution.

Each test can have a different warm-up period, depending on the activity it exercises and potentially also on the software platform (stack) on top of which it executes. The warm-up period is typically determined manually by analysis of data from a very long run, which makes periodic re-evaluation due to changes in the underlying software stack costly.
 
A performance test scheduling would benefit from an automated process that would perform warm-up profiling of test cases added to the system. Even if the initial warm-up profiling took much longer than the typical duration of a test, the savings in subsequent test executions could be substantial. One approach to such warm-up profiling could be based on finding/identifying repetitive patterns in the collected test metrics over time. The automated process would determine the necessary warm-up time and measurement time so as to properly sample the observable variance within a single run. This is a non-trivial challenge, because many benchmarks (or performance) do not exhibit a classic warm-up behavior, and the collected metrics (most often iteration/response time) may oscillate even in steady state~\cite{barrett2016virtual, bondi2016challenges}.

As mentioned earlier, complex long-running performance tests may focus on identifying critical performance regressions in the order of 100\% or more, typically by requiring that the metric of interest is within certain bounds. Since the time scale and the range of such bounds is usually significantly larger than what is common for small-scale tests, it may not be necessary to execute complex long-running tests many times. However, due to the nature of such performance tests, they will still take a long time to execute. The question is then whether all the activity performed by the test is relevant for finding the performance regression, or whether a representative subset would be sufficient. Finding this subset and capturing it in a performance test is non-trivial and challenging, especially if it is to be automated.

\subsubsection{Conclusions}
We assume that the adoption of a DevOps culture will provide both demand and useful information for developers to engage in performance testing. To provide timely feedback based on performance testing to developers, performance tests will have to be scheduled to fit timing and cost constraints, while maximizing the likelihood of discovering a true performance regression. This further relies on the ability to select and prioritize performance tests based on analysis of software changes, utility profile of individual test cases, as well as their membership in equivalence groups of tests producing correlated results. In addition, approaches to determine test-specific warm-up profiles as well as identification of relevant subset of real-world workloads should help in reducing execution times of individual tests, because these are then greatly amplified by the need to test along many dimensions (code changes, configuration changes, hardware platforms, multiple test runs), which leads to testing time explosion. By finding solutions to these problems we should be able to provide a test scheduler with the information necessary to make good scheduling decisions.

%
%


%
%
%
%
%
%
%
%
%
%
%
%
%

\begin{participants}
\participant Alberto Avritzer, Sonatype, USA
\participant Oliver Beck, SAP, Germany
\participant Cor-Paul Bezemer, Queen's University, Canada
\participant Markus Borg, RISE SICS AB, Sweden
\participant Lubom\'ir Bulej, Charles University Prague, Czech Republic
\participant J\"urgen Cito, University of Zurich, Switzerland
\participant Georgiana Copil, TU Vienna, Austria
\participant Robert Heinrich, Karlsruhe Institute of Technology, Germany
\participant Andre van Hoorn, University of Stuttgart, Germany
\participant Vojt\v{e}ch Hork\'{y}, Charles University Prague, Czech Republic
\participant Pooyan Jamshidi, Imperial College London, United Kingdom
\participant Jack Jiang, York University, Canada
\participant Cristian Klein, Umea University, Sweden
\participant Holger Knoche, Christian-Albrechts-Universit\"at zu Kiel, Germany
\participant Philipp Leitner, University of Zurich, Switzerland
\participant Fei Li, Siemens Corporate Technology, Austria
\participant Lucy Ellen Lwakatare, University of Oulu, Finland
\participant Dusica Marijan, Simula, Norway
\participant Claus Pahl, Free University of Bozen-Bolzano, Italy
\participant Stefan Schulte, TU Vienna, Austria
\participant Weiyi Shang, Concordia University, Canada
\participant Catia Trubiani, Gran Sasso Science Institute, Italy
\participant J\"urgen Walter, Julius Maximilians Universit\"at W\"urzburg, Germany
\participant Ingo Weber, Data61, CSIRO, Australia
\participant Johannes Wettinger, University of Stuttgart, Germany
\participant Felix Willnecker, fortiss, Germany
\end{participants}

\enlargethispage{1cm}

\vfill

\begin{figure}[h]
\centering
\includegraphics[width=0.9\textwidth]{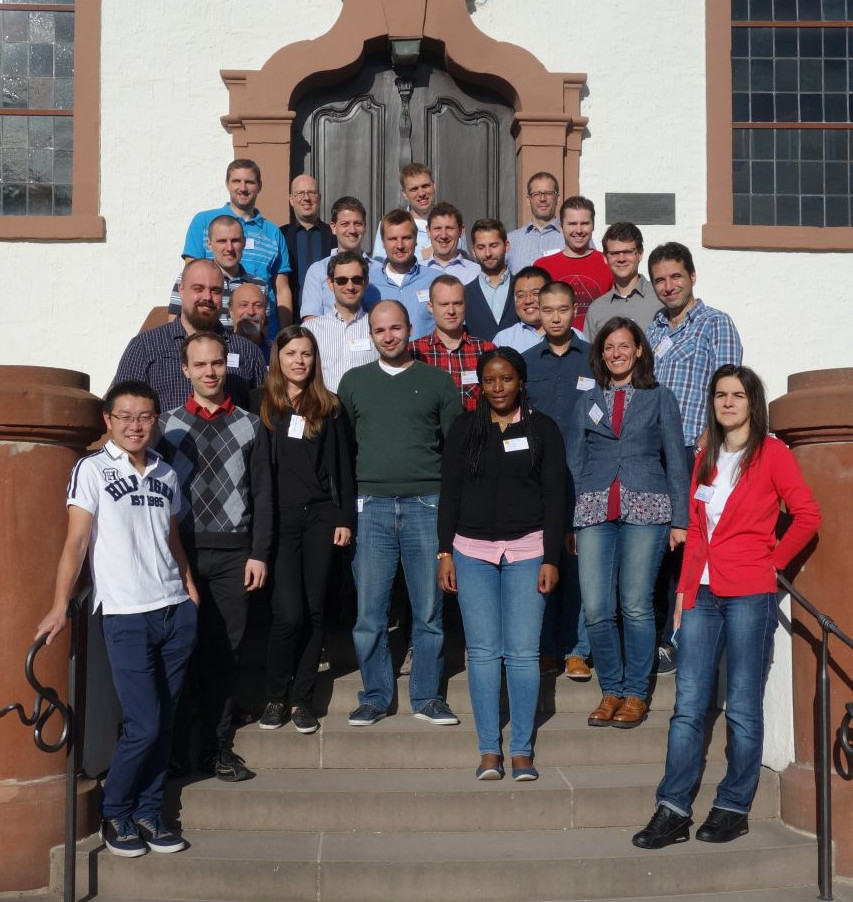}
\end{figure}

\vfill

\newpage

\bibliography{dagrep-sample}

\end{document}